\documentclass[12pt]{article}
\usepackage[utf8]{inputenc}
\usepackage{graphicx,psfrag,epsf}
\usepackage{enumerate}
\usepackage{natbib}

\newcommand{\blind}{0}

\usepackage{hyperref}
\usepackage{lmodern}
\usepackage[fleqn]{amsmath}
\usepackage{amssymb,amsfonts}
\usepackage{color}
\usepackage{fancyvrb}
\usepackage{caption}
\usepackage{latexsym}
\usepackage{transparent}
\usepackage{pifont}
\usepackage{exscale}
\usepackage{dcolumn}
\usepackage{graphicx}
\usepackage{xcolor}
\usepackage{ragged2e}
\usepackage{longtable}
\usepackage{booktabs}
\usepackage{comment}
\usepackage{float}
\usepackage{dsfont}
\usepackage{csquotes}
\usepackage{rotating}
\usepackage{enumitem}
\usepackage{titlesec}
\usepackage{pgf}
\usepackage{bm}
\usepackage{footnote}
\makesavenoteenv{tabular}
\makesavenoteenv{table}
\usepackage{twoopt}
\usepackage{algorithm}
\usepackage{algpseudocode}
\usepackage{multirow}
\usepackage{makecell}
\usepackage{diagbox}
\usepackage{lscape}
\usepackage{pdflscape}
\usepackage{layouts}

\definecolor{light-gray}{gray}{0.95}

\ifdefined\N                                                                
\renewcommand{\N}{\mathds{N}}                                                
\else
  \newcommand{\N}{\mathds{N}}
\fi
\newcommand{\R}{\mathds{R}}                                                 
\ifdefined\C
  \renewcommand{\C}{\mathds{C}}                                             
\else
  \newcommand{\C}{\mathds{C}}
\fi

\def\argmin{\mathop{\sf arg\,min}}                                          
\newcommand{\fp}[2]{\frac{\partial #1}{\partial #2}}                        

\newcommand{\diag}{\operatorname{diag}}                                     

\newcommand{\xv}{\mathbf{x}}													

\renewcommand{\P}{\mathds{P}}                                               

\newcommand{\Xspace}{\mathcal{X}}                                           
\newcommand{\Yspace}{\mathcal{Y}}                                           
\newcommand{\Pxy}{\P_{xy}}                                                  
\newcommand{\xy}{(\mathbf{x}, y)}                                                  
\newcommand{\D}{\mathcal{D}}                                                
\newcommand{\Dset}{\left( \left(\mathbf{x}^{(1)}, y^{(1)}\right), \ldots, \left(\mathbf{x}^{(n)},  y^{(n)}\right)\right)}    
\renewcommand{\xi}[1][i]{\mathbf{x}^{(#1)}}                                          
\newcommand{\yi}[1][i]{y^{(#1)}}                                            
\newcommand{\xj}{\xv_j}                                                       

\newcommand{\fh}{\hat{f}}                                                   
\newcommand{\fxi}{f\left(\mathbf{x}^{(i)}\right)}                                        

\newcommand{\Lxyi}{L\left(\yi, \fxi\right)}                                 

\newcommand{\riske}{\mathcal{R}_{\text{emp}}}                               

\newcommand{\fmh}{\hat{f}^{[m]}}                                            
\newcommand{\fmdh}{\hat{f}^{[m-1]}}                                         
\newcommand{\rmm}{\tilde{\mathbf{r}}^{[m]}}                                                  
\newcommand{\rmi}{\tilde{r}^{[m](i)}}                                               
\renewcommand{\mathbf}{\bm}
\renewcommand{\rmm}{\bm{r}^{[m]}}
\renewcommand{\rmi}{r^{[m](i)}}

\newcommand*{\tran}{{\mkern-1.5mu\mathsf{T}}}

\def\diag{\mathop{\sf diag}}

\renewcommand{\min}{\mathop{\sf min}}
\renewcommand{\max}{\mathop{\sf max}}

\newcommand{\tb}{\bm{\theta}}

\newcommand{\tbih}[1]{\hat{\tb}^{[#1]}}
\newcommand{\tbj}{\bm{\theta}_k}
\newcommand{\tbh}{\hat{\tb}}
\newcommand{\tbmh}{\tbh^{[m]}}
\newcommand{\tbjh}{\hat{\tb}_k}
\newcommand{\bj}{b_k}
\newcommand{\bjm}{b_{k^{[m]}}}

\newcommand{\bd}{d}
\newcommand{\bdj}{{d_k}}

\newcommand{\design}{\bm{Z}}

\newcommand{\xjcat}{\tilde{\xv}_j}

\newcommand{\xjcatvec}{(\tilde{x}^{(1)}_j, \ldots, \tilde{x}^{(n)}_j)^\tran}

\newcommand{\xjcatb}{\bm{z}_j}
\newcommand{\xijcatb}[1][i]{z_j^{(#1)}}

\newcommand{\kj}{\mathbf{ind}_k}
\newcommand{\kij}[1][i]{ind_k^{(#1)}}
\newcommand{\kjvec}{(\kij[1], \ldots, \kij[n])^\tran}

\newcommand{\cmm}{\mathbf{c}^{[m]}}
\newcommand{\cmi}{c^{[m](i)}}

\newcommand{\cmid}{c^{[m-1](i)}}

\newcommand{\bjcorm}{b_{k_\text{cor}^{[m]}}}
\newcommand{\bjcori}[1]{b_{k_\text{cor}^{[#1]}}}
\newcommand{\tbmcorh}[1]{\hat{\bm{\theta}}_\text{cor}^{[#1]}}

\newcommand{\gmd}{g^{[m-1]}}

\newcommand{\pluseq}{\mathrel{+}=}
\newcommand{\pn}{p_\text{noise}}
\newcommand{\snr}{\text{SNR}}
\newcommand{\xij}[1][i]{x_j^{(#1)}}

\def\dtest{\mathcal{D}_\text{val}}
\def\dtrain{\mathcal{D}_\text{train}}

\def\updateCWB{\mathop{\sf updateCWB}}
\def\updateACWB{\mathop{\sf updateACWB}}

\begin{document}


\def\spacingset#1{\renewcommand{\baselinestretch}%
{#1}\small\normalsize} \spacingset{1}

\date{}

\if0\blind
{
  \title{\bf Accelerated Componentwise Gradient Boosting using Efficient Data Representation and Momentum-based Optimization}
  \author{Daniel Schalk, 
    Bernd Bischl 
    and 
    David R{\"u}gamer \\
    Department of Statistics, LMU Munich}
  \maketitle
} \fi


\bigskip
\begin{abstract}

Componentwise boosting (CWB), also known as model-based boosting, is a variant of gradient boosting that builds on additive models as base learners to ensure interpretability. CWB is thus often used in research areas where models are employed as tools to explain relationships in data. One downside of CWB is its computational complexity in terms of memory and runtime. In this paper, we propose two techniques to overcome these issues without losing the properties of CWB: feature discretization of numerical features and incorporating Nesterov momentum into functional gradient descent. As the latter can be prone to early overfitting, we also propose a hybrid approach that prevents a possibly diverging gradient descent routine while ensuring faster convergence. We perform extensive benchmarks on multiple simulated and real-world data sets to demonstrate the improvements in runtime and memory consumption while maintaining state-of-the-art estimation and prediction performance.
\end{abstract}

\noindent%
\vfill
\newpage

\spacingset{1.5}


\section{Introduction}
\label{sec:intro}


\textit{Model-based} or \textit{componentwise boosting}~\citep[CWB;][]{buhlmann2003boosting} applies gradient boosting~\citep{freund1996experiments} to statistical models by sequentially adding pre-defined components to the model. These components are so-called base learners of one or multiple features. If interpretable base learners are used (e.g., univariate splines), the full CWB model remains interpretable and allows for the direct assessment of estimated partial feature effects. Further advantages of CWB are its applicability in high-dimensional feature spaces (\enquote{$p \gg n$ situations}), its inherent variable selection, and unbiased feature selection~\citep{buhlmann2003boosting,hofner2011framework}. It is also possible to derive inference properties of boosted estimators to quantify uncertainty using post-selection inference procedures~\citep{rugamer2019inference}. These properties make CWB a powerful method at the intersection of (explainable) statistical modelling and (black-box prediction) machine learning. For this reason, CWB is frequently used in medical research, e.g., for oral cancer prediction~\citep{saintigny2011gene}, detection of synchronization in bioelectrical signals~\citep{rugamer2016boosting}, or classifying pain syndromes~\citep{liew2020classifyingneck}. In contrast, many other gradient boosting methods such as XGBoost~\citep{chen2016xgboost} solely focus on predictive performance and (mainly) use tree-based base learners with higher-order interactions. As a consequence, these procedures require techniques from interpretable machine learning~\citep[see, e.g.,][]{molnar2020interpretable} to explain their resulting predictions. 

Various versions of the original CWB algorithm have been developed, e.g., CWB for functional data~\citep{JSSv094i10}, boosting location, scale and shape models~\citep{hofner2014gamboostlss}, or probing for sparse and fast variable selection~\citep{thomas2017probing}. CWB's computational complexity in terms of memory and runtime is, however, a downside often encountered in practice. The high consumption of RAM of the current state-of-the-art implementation \texttt{mboost}~\citep{mboost} can considerably exceed the capacity of modern workstations. This makes CWB less attractive or even infeasible for medium- to large-scale applications. In this paper, we focus on two internal structures of CWB and propose improvements to mitigate these problems.

\textbf{Our contributions and related literature}. Our first contribution (Section~\ref{subsec:binning}) is a novel CWB modification to reduce memory consumption in large data situations. To the best of our knowledge, we are the first to derive the complexity of CWB and also the first to suggest improvements to reduce computational costs. Based on a recently proposed idea to fit generalized additive models (GAMs) on large data sets~\citep{li2020faster}, we describe adaptions of matrix operations for CWB to operate on discretized features. We refer to this approach as binning. Binning is a discretization technique for numerical features and can drastically reduce runtime and memory consumption when fitting GAMs, especially when coupled with specialized matrix operations. In contrast to~\citet{li2020faster}, we use binning within each base learner rather than processing the model matrix of all features. This makes the use of binning particularly beneficial for CWB. 

In Section~\ref{subsec:optim}, we also adapt a novel optimization technique called Accelerated Gradient Boosting Machine~\citep[AGBM;][]{lu2020accelerating}. AGBM allows incorporation of Nesterov momentum for gradient boosting in the function space. Based on AGBM, we propose a new variant of CWB for faster convergence while still preserving CWB's interpretability. The adaption perfectly fits to the general optimization scheme of CWB, which is also based on functional gradient descent. However, the acceleration based on Nesterov momentum is known to diverge in certain cases~\citep{wang2020scheduled}. We thus also propose a refinement of AGBM in our algorithm Hybrid CWB (HCWB) to overcome premature divergence of the gradient descent routine. 

These proposed adaptions can be applied to CWB independently of each other and are both implemented in the software package \texttt{compboost}~\citep{schalk2018compboost}. In a simulation study in Section~\ref{sec:experiments}, we study their effect both separately and when combined. Finally, we conduct real-world benchmark experiments in Section~\ref{sec:experiments} to compare our proposed CWB variants with vanilla CWB, the CWB implementation in \texttt{mboost}, XGBoost, and a recently published method called Explainable Boosting Machine (EBM) within the \texttt{interpretML} framework~\citep{nori2019interpretml}.


\section{Componentwise Gradient Boosting}
\label{sec:cwb-basics}

This section introduces the main concepts and properties of CWB as well as the notation used throughout this article.

\subsection{Terminology}
\label{subsec:terms}

Consider a $p$-dimensional feature space $\Xspace = (\Xspace_1 \times \hdots \times \Xspace_p)$ and a {target space} $\Yspace$. We assume an unknown {functional relationship} $f$ between $\Xspace$ and $\Yspace$. ML algorithms try to learn this relationship using a {training data set} $\D = \Dset$ with $n$ observations drawn independently from an unknown {probability distribution} $\Pxy$ on the joint space $\Xspace \times \Yspace$. Let $\fh$ be the estimated model fitted on the training data to approximate $f$ and $P = \{1, \hdots, p\}$ an {index set} for all $p$ features. The vector $\xj = (\xij[1], \ldots, \xij[n])^\tran \in \Xspace$ refers to the $j$th feature. $\xv$ and $y$ are arbitrary members of $\Xspace$ and $\Yspace$, respectively. Given $f$ and a loss function $L: \Yspace \times \R \to \R^+_0$, we assess the prediction quality of the model fit on the data set $\D$ using the \textit{empirical risk} $\riske(f) = |\D|^{-1}\sum_{\xy \in \mathcal{D}}L(y, f(\xv))$.

\subsection{Base Learner}
\label{subsec:blearner}

A base learner $b_k : \Xspace \to \R$ is used to model the contribution of one or multiple features to the estimated model $\fh$. Possible base learners range from simple models like a linear model on one feature to complex models such as deep decision trees including many features. While the presented adaptions for CWB work also for multivariate base learners such as tensor product splines, we restrict ourselves to univariate functions $\bj(\xv, \tb) = g_k(x)^\tran\tb, \tb\in\R^{\bdj}$, for demonstration purposes. The function $g_k : \R \to \R^\bdj$ is a generic representation for modeling alternatives such as linear effects, categorical effects or smooth effects. For smooth effects, numerical features are transformed using a basis representation $g_k(x) = (B_{k,1}(x), \dots, B_{k,\bdj}(x))^\tran$ with $\bdj$ basis functions, e.g., via univariate penalized B-splines~\citep[P-Splines;][]{eilers1996flexible}. This generic representation defines a \textit{design matrix} $\design_k = (g_k(\xij[1]), \ldots, g_k(\xij[n]))^\tran\in\R^{n\times \bdj}$ given by a feature vector $\xj\in\R^n$. Note that the base learner $\bj$ implicitly selects the feature(s) on which it operates without explicitly denoting the feature(s). In this paper with univariate base learners, exactly one feature is passed to the generic representation $g_k$. An important property induced by the choice of a linear base learner is that two base learners of the same type $b_k(\xv, \tb)$ and $b_k(\xv, \tb^\ast)$ sum up to one base learner of the same type:
\begin{equation}\label{eq:bl-additivity}
    b_k(\xv,\tb) + b_k(\xv,\tb^\ast) = b_k(\xv, \tb + \tb^\ast).
\end{equation}
During boosting, CWB selects its next component from a pre-defined pool of base-learners $\mathcal{B} = \{\bj\}_{k \in \{1, \ldots, K\}}$, where in our case of univariate $\bj$, $K$ often equals the number of features $p$.

\subsection{Componentwise Boosting}
\label{subsec:cboosting}

CWB estimates $\fh$ using an iterative steepest descent minimization in function space. 
$\fmh$ denotes the prediction model after $m$ iterations. In each step, the \textit{pseudo residuals} $\rmi = -\fp{\Lxyi}{f(\xi)}\left.\vphantom{f_{(i}^{(i}}\right|_{f = \fmdh},\  i \in \{1, \dots, n\}$ indicate a functional gradient, evaluated at training data points, where changing the outputs of
$\fmh$ reduces the overall loss of our current model, w.r.t. to the given loss $L$ the most. 
CWB initializes $\fmh$ with a loss-optimal constant model -- also called an offset or intercept. In the $m$th iteration, all base learners in $\mathcal{B}$ are fitted against $\rmm$ via L2-loss, and the best candidate is selected to additively update $f^{[m]}$, controlled by a \textit{learning rate} $\nu$. This procedure is repeated $M$ times or until a convergence criterion is met (e.g., using early stopping as described in Section~\ref{subsec:acwb-restart}). The details of CWB are given in Algorithm~1 in Appendix A.1.

Due to property~\eqref{eq:bl-additivity} of linear base learners and the additive model update of CWB, the estimated parameter of each base learner can be summed up after $M$ iterations, and each aggregated parameter $\tbjh = \sum_{m=1}^M\sum_{k=1}^K\mathds{1}_{\{k=k^{[m]}\}}\tbh^{[m]}$ with its base representation $g_k$ can be interpreted as a partial effect of the feature modeled by base learner $\bj$. 

\subsection{Computational Complexity of CWB}\label{subsec:cwb-complexity}

The computational complexity of CWB is directly related to the computational complexity of fitting each base learner. All linear base learners $b_1, \ldots, b_K$ must solve a system of linear equations $\design_k^\tran\design_k\tbj = \design_k^\tran \rmm$. For simplicity, assume that all base learners have the same number of parameters, i.e., $\bdj = \bd$. Such systems are usually Cholesky decomposed with $\design_k^\tran\design_k = \bm{L}_k\bm{L}_k^\tran$, and complexity $\mathcal{O}(\bd^3)$. 
Taking into account additional matrix operations $\design_k^\tran \design_k$ with $\bd^2n$, $\design_k^\tran \rmm$ with $\bd n$, and forward/backward solving with $(\bd^2 - \bd)/2$ operations, this yields a complexity of $\mathcal{O}(\bd^2n + \bd^3)$. 
When applied to all $K$ base learners in $M$ iterations, this yields $\mathcal{O}(MK(\bd^2n + \bd^3))$ (neglecting operations such as the calculations of pseudo residuals $\rmm$ and sum of squared errors $\text{SSE}_k$, or finding the best base learner $k^{[m]}$). 

The above can be accelerated considerably by pre-calculating the (expensive) Cholesky decomposition once for every base learner before boosting. This reduces the computational complexity of CWB to $\mathcal{O}(K(\bd^2n + \bd^3))$. When taking also the forward/backward solving and calculation of $\design_k^\tran \rmm$ as operations per iteration into account, the complexity when caching heavy operations reduces to $\mathcal{O}(K(\bd^2n + \bd^3) + MK(\bd^2 + dn))$.

\section{A more Efficient Componentwise Boosting}

\subsection{Binning}
\label{subsec:binning}

In order to make CWB feasible for large data sets and reduce computational resources in general, we propose to combine CWB with binning. While the primary goal of binning is to reduce the memory consumption by representing numerical feature via discretization, this method can also accelerate the model fit.

\subsubsection{Discretizing numerical Features}

Binning discretizes a numerical feature into a smaller number of design point values.
Usually, binned values are constructed as an equally spaced grid \cite{lang2014multilevel} $\xijcatb = \min(\xj) + (i-1)/(n^\ast - 1) (\max(\xj) - \min(\xj))$, $i = 1, \dots, n^\ast$ and then replace each value $\xij$ with its closest design point $\xijcatb$. The number of design points $n^\ast$ can be chosen arbitrarily. \cite{wood2017generalized} argue that the trade-off between data size and statistical error due to imprecise feature values is most adequate for $n^\ast = \sqrt{n}$. 

\textbf{Design points and index vector}. Instead of storing the discretized feature vector $\xjcat = \xjcatvec\in\R^n$, it is sufficient to store the $n^\ast$ values of $\xjcatb$ as well as an additional index vector $\kj = \kjvec$ (i.e., the assignment of each discretized feature value to its bin $\tilde{x}_k^{(i)} = \xijcatb[\kij]$). The index vector is calculated by setting $\kij = 1$ if  $\xij \in [\xijcatb[0];\xijcatb[1] + m_2]$ and $\kij = l$ if $\xij \in (\xijcatb[l-1] - m_1; \xijcatb[l] + m_2 ]$, where $m_1 = (\xijcatb[l] - \xijcatb[l-1]) / 2$ and $m_2 = (\xijcatb[l+1] - \xijcatb[l])/2$ is half the distance of a design point to its left/right neighbor. Hence, binning can be seen as a hash map where the index vector $\kj$ is the hash function, the design points $\xjcatb$ are the keys, and $\xj$ are the entries.

Binning reduces the amount of required storage for variables. Instead of storing the $n\times \bdj$ matrix $\design_k$, a reduced $n^\ast \times \bdj$ matrix $\tilde{\design}_k^b = (g_k(\xijcatb[1]), \ldots, g_k(\xijcatb[n^\ast]))^\tran$ based on bins $\xjcatb$ is stored. $\bm{k}_k$ is used to assign the $i$th row $\tilde{\design}_k^b(i)=g_k(\xijcatb[i])^\tran$, $i = 1, \ldots, n^\ast$ to the $i$th row $\tilde{\design}_k(i) = g_k(\tilde{x}_k^{(i)})^\tran$, $i = 1, \dots, n$ of the discretized feature $\xjcat$ by $\tilde{\design}_k(i) = \tilde{\design}_k^b(\kij)$. Note that the same lookup can also be applied to categorical features without additional binning.

An analogy to binning are sparse data matrices, a widely used data representation. Similar to binning, sparse data matrices choose another representation by storing the row and column index of only the non-zero elements and the corresponding values \citep[see, e.g.,][]{duff1989sparse}. Using sparse matrices has two major advantages. First, this approach incurs reduced memory load, and second, optimized algorithms can be used to calculate matrix operations much faster. A specific example where sparse matrices are used in the context of CWB is to store the base representation $\design_k$ of P-spline base learners, since $\design_k$ contains mainly zeros. The fitting process is also accelerated, as $\design_k^\tran\rmm$ is calculated for just the non-zero elements.

\subsubsection{Matrix multiplications on binned Features}\label{subsubsec:binning-algo}

When fitting a penalized base learner in CWB, the two matrix operations $\design_k^\tran\bm{W}_k\design_k + \lambda \bm{D}_k$ and $\design_k^\tran\bm{W}_k\rmm$ are required. We assume the weight matrix to be diagonal $\bm{W}_k = \diag(\bm{w}_k)$ with elements $\bm{w}_k = (w_k^{(i)}, \dots, w_k^{(n)})^\tran$ and set $\bm{w}_k$ to a vector of ones if no weights are used, i.e., $\bm{W}_k \equiv \bm{I}_n$. While $\lambda\bm{D}_k$ is not affected by binning, $\design_k^\tran\bm{W}_k\design_k$ and $\design_k^\tran\bm{W}_k\rmm$ can be computed more efficiently on the binned design matrix. Algorithm~\ref{algo:binning-mat-mult} describes corresponding matrix operations using index vector $\bm{k}_k$ and reduced design matrix $\tilde{\design}_k^b$. 

\spacingset{1}

\begin{algorithm}[H]
\caption{Calculation of $
\tilde{\design}_k^\tran\bm{W}_k\tilde{\design}_k$ and $\tilde{\design}_k^\tran\bm{W}_k\rmm$ on binned design matrix $\tilde{\design}^b_k$ and weight matrix $\bm{W}_k = \mathsf{diag}(\bm{w}_k)$.}\label{algo:binning-mat-mult}
\footnotesize
\vspace{0.15cm}
\hspace*{\algorithmicindent} \textbf{Input:} Design matrix $\tilde{\design}_k^{b}$, weight vector $\bm{w}_k$, pseuro residuals $\rmm$, and index vector $\kj$\\
\hspace*{\algorithmicindent} \textbf{Output:} $\tilde{\design}_k^\tran\bm{W}_k\tilde{\design}_k$ and $\tilde{\design}_k^\tran\bm{W}_k\rmm$
\vspace{0.15cm}
\hrule
\vspace{0.15cm}
\noindent\begin{minipage}{\textwidth}
    \begin{minipage}{0.50\textwidth}
        \begin{algorithmic}[1]
            \Procedure{binMatMat}{$\tilde{\design}_k^b,\bm{W}_k, \kj, \bm{w}_k$}
            \State Initialize with zero matrix $\bm{U} = \bm{0}_{\bdj\times n^\ast}$
            \State \textbf{for} $i \in \{1, \dots, n\}$ \textbf{do} 
            \State \hspace{1cm}$\bm{U}(\kij) \pluseq w_{i,j}\tilde{\design}_k^b(k_k^{(i)})$
            \State \textbf{return} $\bm{U}\tilde{\design}_k^b$
            \EndProcedure
        \end{algorithmic}
    \end{minipage}
    \hfill
    \begin{minipage}{0.48\textwidth}
        \begin{algorithmic}[1]
            \Procedure{binMatVec}{$\tilde{\design}_k^b, \rmm, \kj, \bm{w}_k$}
            \State Initialize with zero vector $\bm{u} = \bm{0}_{n^\ast}$
            \State \textbf{for} $i \in \{1, \dots, n\}$ \textbf{do}
            \State \hspace{1cm} $u^{(\kij)} \pluseq w_k^{(i)}\rmi$
            \State \textbf{return} $\tilde{\design}_k^{b\tran}\bm{u}$
            \EndProcedure
        \end{algorithmic}
    \end{minipage}%
\end{minipage}
\vspace{0.3cm}
\end{algorithm}

\spacingset{1.5}

The matrix $\bm{U}$ and vector $\bm{u}$ act as intermediate results and are used for the final matrix-matrix and matrix-vector operation on the reduced design matrix $\tilde{\design}^b$. Due to the discretization, the number of matrix product operations reduces from $\mathcal{O}(n\bd^2)$ to $\mathcal{O}(n^\ast \bd^2 + n)$ when using a diagonal weight matrix \citep{li2020faster}.

CWB applies binning on a base learner level. Therefore, each base learner that uses binning individually calculates the bin values $\xjcatb$, the reduced design matrix $\tilde{\design}_k^b$, and the index vector $\kj$. The \textsc{binMatMat} in Algorithm~\ref{algo:binning-mat-mult} is first used in CWB when calculating the Cholesky decomposition $\bm{L}$ of $\tilde{\design}_k^\tran\bm{W}_k\tilde{\design}_k$ and then caches results for later usage. To calculate $\tbmh$ in each iteration, \textsc{binMatVec} in Algorithm~\ref{algo:binning-mat-mult} is used.

\subsubsection{Computational Complexity when applied in CWB}\label{subsubsec:binning-comp-complexity}

Using Algorithm~\ref{algo:binning-mat-mult}, the number of operations for calculating $\tilde{\design}_k^\tran\bm{W}_k\tilde{\design}_k$ reduces from  $K(\bd^2n + d^3)$ to $K(n^\ast\bd^2 + n + d^2)$. In comparison to a routine without binning, this is a reduction in operations if $n^\ast < n(\bd^2 - 1)/\bd^2$ and $\bd > 1$. A similar result holds during the fitting process when applying {binMatVec} of Algorithm~\ref{algo:binning-mat-mult}. Here, binning requires $n + \bd n^\ast$ operations, whereas a calculation with dense matrices requires $\bd n$ operations. This is a reduction in operations if $n^\ast < n(\bd - 1)/\bd$ and $\bd > 1$. Applying this for all $K$ base learners in each of the $M$ iterations yields a complexity of $\mathcal{O}(MK(\bd^2 + dn^\ast + n)$. All in all, we obtain a computational complexity of $\mathcal{O}(K(\bd^2n^\ast  + n + \bd^3) + MK(\bd^2 + \bd n^\ast + n))$ instead of $\mathcal{O}(K(\bd^2n + \bd^3) + MK(\bd^2 + \bd n))$. For the important case of P-spline base learners with $\bd = 20$ as a typical choice of basis dimension and $n^\ast = \sqrt{n}$, the conditions for a reduction in operations are fulfilled (see also Section~\ref{sec:experiments} for the effect of binning in practice).

\subsection{Accelerating the Fitting Process of CWB}
\label{subsec:optim}

In risk minimization, standard gradient descent can evolve slowly if the problem is ill-conditioned~\citep{ruder2016overview}. To reduce this problem, momentum adds a fraction of the previous gradient to the update step for an accelerated fitting procedure~\citep{qian1999momentum}. A further extension is Nesterov accelerated gradient \citep[NAG;][]{NESTEROV1983}, also known as Nesterov momentum. NAG performs a look ahead on  what the gradient descent step is doing and adjusts the update step to improve convergence. To incorporate NAG into CWB, we follow \citet{lu2020accelerating} and introduce a second base learner $\bjcorm$ that is fitted to the so-called \textit{error-corrected pseudo residuals} $\cmi = \rmi + \frac{m}{m+1}(\cmid - \bjcori{m-1} (\xi, \tbmcorh{m}),\ i \in \{1, \dots, n\}$. Due to the recursive definition, the error-corrected pseudo residual $\cmi$ at iteration $m$ contains information of all previous pseudo residuals. 
The sequence $\{\bjcori{1}, \dots, \bjcori{m}\}$ of base learners then accumulates to the \textit{momentum model} $h^{[m]} = h^{[m-1]} + \eta_m\bjcorm$, where $\eta_m = \beta\gamma\vartheta_m^{-1}$ is the learning rate of the momentum model. This learning rate contains the momentum parameter $\gamma \in\mathbb{R}_+$ and a sequence $\vartheta_m = 2 / (m+1)$, which is later used to combine the \textit{primary model} $f$ and momentum model $h$. In contrast to standard CWB, the pseudo residuals $\rmi = -\fp{\Lxyi}{g(\xi)}\left.\vphantom{f_{(i}^{(i}}\right|_{g = \gmd},\ i \in \{1, \dots, n\} $ are calculated as the gradient w.r.t. a convex combination $g^{[m]}=(1-\vartheta_m)f^{[m]} + \vartheta_m h^{[m]}$ of the primary model $f^{[m]}$ and the momentum model $h^{[m]}$. The primary model $f^{[m]} = g^{[m-1]} + \beta \bjm$ is calculated by adding the new component $\bjm$ to the combined model $g^{[m]}$. Algorithm~\ref{algo:acc-comp-boosting} summarizes the \textit{accelerated CWB} (ACWB) algorithm. For simplicity, the loop to select the optimal base learner (lines 5 -- 8 of Algorithm~1 Appendix~A.1) is summarized as procedure ${\sf findBestBaselearner}(\mathbf{r}, \mathcal{B})$, which returns a tuple $(\tbmh, k^{[m]})$ of the estimated parameters $\tbmh$ and the index $k^{[m]}$ of the best base learner from set $\mathcal{B}$ of base learners.

\spacingset{1}

\begin{algorithm}[h]
 \caption{ACWB algorithm with input and output.}\label{algo:acc-comp-boosting}
\vspace{0.15cm}
\hspace*{\algorithmicindent} \textbf{Input} Train data $\D$, learning rate $\nu$, momentum parameter $\gamma$, number of boosting\\
\hspace*{\algorithmicindent} \phantom{\textbf{Input} }iterations $M$, loss function $L$, set of base learner $\mathcal{B}$\\
\hspace*{\algorithmicindent} \textbf{Output} Prediction model $\hat{f}^{[M]}$ defined by fitted $\tbih{1}, \ldots, \tbih{M}$ and $\tbmcorh{1}, \ldots, \tbmcorh{M}$\vspace{0.15cm}
\hrule
\begin{algorithmic}[1]
\Procedure{ACWB}{$\D,\nu,\gamma,L,\mathcal{B}$}
    \State Initialize $\fh^{[0]}(\xv) = \argmin\limits_{c\in\R} \riske(c,\D)$;\ \ $\hat{h}^{[0]}(\xv) = \fh^{[0]}(\xv)$;\ \ $\hat{g}^{[0]}(\xv) = \fh^{[0]}(\xv)$;

    \For{$m \in \{1, \dots, M\}$}
        \State $\vartheta_m =\frac{2}{m+1}$ \label{algo:acwb-start}
        \State $\hat{g}^{[m]}(\xv) = (1 - \vartheta_m) \fh^{[m-1]}(\xv) + \vartheta_m \hat{h}^{[m-1]}(\xv)$
        \State $\rmi = -\left.\fp{\Lxyi}{g(\xi)}\right|_{g = \hat{g}^{[m]}},\ \ \forall i \in \{1, \dots, n\}$
        \State $(\tbmh, k^{[m]}) = {\sf findBestBaselearner}(\rmm, \mathcal{B})$
        \State $\fmh(\xv) = \hat{g}^{[m]}(\xv) + \nu \bjm(\xv, \tbmh)$
        \If{$m > 1$}
            \State $\cmi = \rmi + \frac{m}{m+1}\left(\cmid - \bjcori{m-1} (\xv, \tbmcorh{m}\right),\ \ \forall i \in \{1, \dots, n\}$
        \Else
            \State $\cmm = \rmm$
        \EndIf
        \State $(\tbmcorh{m}, k^{[m]}_\text{cor}) = {\sf findBestBaselearner}(\cmm, \mathcal{B})$ \label{algo:acc-comp-boosting:bl2}
        \State $\eta_m = \gamma\nu\vartheta_m^{-1}$
        \State $\hat{h}^{[m]}(\xv) = \hat{h}^{[m-1]}(\xv) + \eta_m   \bjcorm(\xv,\tbmcorh{m})$\label{algo:acwb-end}
    \EndFor
    \State \textbf{return} $\fh = \hat{f}^{[M]}$
\EndProcedure
\end{algorithmic}
\end{algorithm}
\spacingset{1.5}

\subsubsection{Retaining CWB Properties}

ACWB fits a second base learner $\bjcorm$ in order to accelerate the fitting process with the momentum model $h$. In addition to the \textit{fitting trace} for the primary model $\Theta_f = \{\tbih{1}, \ldots, \tbih{M}\}$, another fitting trace $\Theta_h = \{\tbmcorh{1}, \ldots, \tbmcorh{M}\}$ is also stored for the momentum model. Consequently, ACWB contains twice as many base learners as CWB after $m$ iterations. Despite the more complex fitting routine, the parameters for ACWB can still be additively updated. Suppressing the iteration $m$ for readability, we assign $(1-\vartheta_m)\tbh_f + \vartheta_m \tbh_h$ as the current $\tbh_g$, update $\tbh_f$ to $\tbh_g + \beta(0 \cdots \tbh^{[m]\tran} \cdots 0)^\tran$, and update $\tbh_h$ to $\tbh_h + \eta_m(0 \cdots \tbh^{[m]\tran}_{\text{cor}}\cdots 0)^\tran$. The initial parameters $\tbh_f$ and $\tbh_h$ for $m=0$ are set to zero. An additive aggregation of parameters is important to retain interpretable additive partial effects and allows for much faster predictions.

\subsubsection{Computational Complexity of ACWB}\label{subsec:complexity-acwb}

As derived in Section~\ref{subsec:cwb-complexity}, the complexity of CWB is $\mathcal{O}(K(\bd^2n + \bd^3) + MK(\bd^2 + dn))$ and hence scales linearly in the number of rows $n$ and number of base learners $K$. By fitting a second base learner in each iteration, the complexity for ACWB during the fitting process is doubled compared to CWB, while costs for expensive pre-calculation steps do not change. This results in a complexity of $\mathcal{O}(K(\bd^2n + \bd^3) + 2MK(\bd^2 + dn))$.

\subsubsection{A Hybrid CWB Approach}\label{subsec:acwb-restart}
\spacingset{1}
\begin{algorithm}
\caption{HCWB as a combination of ACWB and CWB. $\riske(f|\dtest)$ denotes the empirical risk calculated on validation data $\dtest$.}\label{algo:acwb-restart}
\vspace{0.15cm}
\hspace*{\algorithmicindent} \textbf{Input} Train data $\D$, learning rate $\nu$, momentum parameter $\gamma$, number of boosting\\
\hspace*{\algorithmicindent} \phantom{\textbf{Input} }iterations $M$, patience parameter $pat_0$, loss function $L$, set of base learner $\mathcal{B}$\\
\hspace*{\algorithmicindent} \textbf{Output} Prediction model $\hat{f}^{[M]}$ defined by parameters $\tbih{1}, \ldots, \tbih{M}$ and $\tbmcorh{1}, \ldots, \tbmcorh{M}$\vspace{0.15cm}
\hrule
\begin{algorithmic}[1]
\Procedure{HCWB}{$\D,\nu,\gamma,pat_0,L,\mathcal{B}$}
    \State $m = 0$
    \For{$m = 1, \ldots, M$}
        \If {$pat_0 < pat$}
            \State $\fmh = \updateACWB(\fmdh, \dtrain, \nu, \gamma,L\mathcal{B})$
            \State \textbf{if} $\riske(\fmh| \dtest) > \riske(\fmdh| \dtest)$ \ \ \textbf{then}\ \ $pat_0 \pluseq 1$ \ \ \textbf{else}\ \ $pat_0 = 0$
        \Else
            \State $\fmh = \updateCWB(\fmdh, \D, \nu,L,\mathcal{B})$
        \EndIf
    \EndFor
    \State \textbf{return} $\hat{f} = \hat{f}^{[m]}$
\EndProcedure
\end{algorithmic}
\end{algorithm}
\spacingset{1.5}
\noindent It is well known that excessive training boosting can lead to overfitting \citep[see, e.g.,]{grove1998boosting, jiang2004process}. This can be controlled by stopping the algorithm early, which works irrespective of using gradient descent with or without NAG. However, as shown recently by \citet[Theorem 4]{wang2020scheduled}, gradient methods with NAG might diverge for noisy convex problems. Whereas inexact gradients induced by noise terms influence the convergence rates only by an additional constant when using no acceleration, NAG potentially diverges with a rate that increases with the number of iterations. To overcome this issue, we propose a hybrid approach by starting the fitting process with ACWB on data $\dtrain \subset \D$ until it is stopped early using a validation set $\dtest$ ($\dtest \cap \dtrain = \emptyset$ and $\dtest \cup \dtrain = \D$). Thereafter, the training is continued to fine-tune the predictions using the classical CWB on all available data $\D$ until $M$ iterations are reached (or, alternatively, using again a train-validation split and early stopping). Algorithm~\ref{algo:acwb-restart} describes this hybrid approach. The update routines $\updateCWB$ and $\updateACWB$ describe the fitting component of the respective algorithms (CWB; Algorithm~1 Appendix~A.1 lines~4 --~10, ACWB; Algorithm~\ref{algo:acc-comp-boosting} lines~\ref{algo:acwb-start} --~\ref{algo:acwb-end}). 
\begin{figure}
    \centering
    \includegraphics{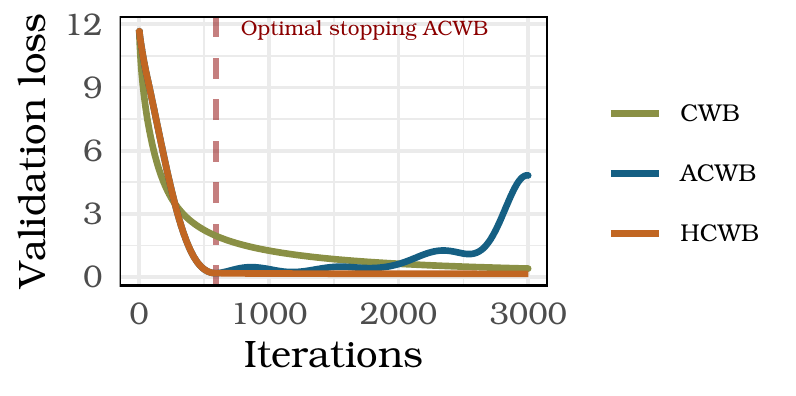}
    \caption{\footnotesize Exemplary course of the empirical risk for different CWB variations on a test data. The red dashed line indicates the optimal stopping iteration found by ACWB. All three methods are trained for 3000 iterations.}
    \label{fig:optim-empr-risk}
\end{figure}
The example shown in Figure~\ref{fig:optim-empr-risk} depicts the behavior of our proposed adaption. After early stopping ACWB, we can continue to further decrease the validation error, whereas continuing training with acceleration (ACWB) increases the validation error. Fine-tuning ACWB thus improves the performance of ACWB while requiring less iterations compared to CWB with a similar validation error. In Section~\ref{subsec:bm-sim-data}, we investigate computational advantages of the hybrid approach in terms of runtime, memory consumption, and estimation properties.

\section{Experiments}
\label{sec:experiments}

We study the efficacy of the proposed improvements on simulated and real-world data (cf. Appendix~B.2). All CWB variants (CWB, ACWB, and HCWB) are either denoted with \emph{(nb)} if no binning is applied, or \emph{(b)} if binning is used. The hyperparameters (HPs) of our proposed algorithm are the amount of binning, the momentum rate $\gamma$, and the patience parameter $pat_0.$ Further HPs for all CWB variants (including vanilla CWB) are the learning rate $\nu$, the degrees of freedom df, and the number of boosting iterations $M$. As simulation studies and real-world applications have different purposes, we will define separate HP tuning schemes for both but set the number of bins to $n^\ast = 2$, as suggested by \cite{wood2017generalized}, and the patience parameter to $pat_0 = 5$ for all experiments. HPs defining the flexibility of base learners such as number of knots or the degree of basis functions for P-Spline base learners are set to 20 knots and degree 3, respectively.

In Section~\ref{subsec:bm-sim-data}, we first use simulated data to investigate the following hypotheses:
\begin{itemize}
    \item[\textbf{H1}]\textbf{(memory and runtime efficiency of binning):} 
        Compared to CWB (nb), binning reduces memory consumption and runtime.
        
    \item[\textbf{H2}]\textbf{(partial effects of binning):} 
        Compared to CWB (nb), using CWB (b) does not deteriorate the estimation quality of partial effects. 
        
    \item[\textbf{H3}]\textbf{(iterations and partial effects of accelerated methods):}
        Compared to CWB (nb), ACWB (nb) and HCWB (nb) require fewer iterations to achieve almost identical partial effect estimates.
        
    \item[\textbf{H4}]\textbf{(scaling behaviour of ACWB/HCWB):} Empirical runtimes of binning and ACWB follow the previously derived complexity conditions and thus scale efficiently with an increasing number of observations and base learners. 
        
\end{itemize}
In addition to the simulation study, we conduct a comparison of CWB and HCWB on real-world data sets in Section~\ref{subsec:bm-real-data}. Here, we investigate the following experimental questions:
\begin{itemize}
    \item[\textbf{EQ1}]\textbf{(implementation comparison with the state-of-the-art software \texttt{mboost}):} Compared to \texttt{mboost}, our implementation \texttt{compboost} is less time-consuming during model training.

    \item[\textbf{EQ2}]\textbf{(algorithmic comparison of accelerated methods):} 
        When compared to vanilla CWB (nb), the prediction performance of the five proposed CWB variants (CWB (b), ACWB/HCWB (b)/(nb)) does not deteriorate while yielding faster runtimes.

    \item[\textbf{EQ3}]\textbf{(comparison with state-of-the-art algorithms)} 
        The prediction performance of ACWB (b) and HCWB (b) is competitive with state-of-the-art boosting algorithms while yielding notably faster runtimes. 
\end{itemize}
All simulations and benchmarks are conducted using \texttt{R} \citep{rmanual} version 3.6.3 on three identical servers with 32 2.60 GHz Intel(R) Xeon(R) CPU e5-2650 v2 processors. Reproducibility of benchmark and simulation results is ensured by providing a Docker image with pre-installed software and benchmark code. References to GitHub repositories containing our \texttt{compboost} software and the raw benchmark source code with results are referenced in the Supplementary Material.

\subsection{Numerical Experiments on Simulated Data}\label{subsec:bm-sim-data}

\subsubsection{Simulation Setup}
\label{subsubsec:sim-setup}

We define the number of informative and noise features as $p$ and $p_\text{noise}$, respectively. Numerical features $\xj$ are simulated by drawing the minimum value $x_{j,\text{min}}$ from a uniform distribution $U[0,100]$ and the maximal value $x_{j,\text{max}} = x_{j,\text{min}} + \rho_k$, where $\rho_k$ also follows a uniform distribution $U[0,100]$. The $n$ feature values $\xj$ are simulated i.i.d. from $U[x_{j,\text{min}}, x_{j,\text{max}}]$. All feature effects are simulated as non-linear effects using splines. The spline basis $\design_k$ for the simulation is created using feature $\xj$ with splines of order 4 and 10 inner knots. To obtain unique splines for each $\xj$, we sample $\tau_k\sim\mathcal{N}_{10}(0,9)$ and define the $j$th feature effect as $\bm{\eta}_k= \bm{Z}_k\bm{\tau}_k$. Each of the $\pn$ noise features $\dot{\bm{x}}_{j}$ are drawn from a standard normal distribution. The final data set is given as $\D = \{(x_1^{(i)}, \dots, x_p^{(i)}, \dot{x}_1^{(1)}, \dots, \dot{x}_{\pn}^{(i)}, \yi )\ | \ i = 1, \dots, n\}$ with target vector $\bm{y} \sim \mathcal{N}_n(\bm{\eta}, \diag_n{(\hat{\sigma}^2(\bm{\eta}) / \snr^2}))$, where SNR is the \textit{signal-to-noise} ratio, $\bm{\eta} = \sum_{j=1}^p\bm{\eta}_k$ and $\hat{\sigma}(\bm{\eta})$ the sample variance of $\bm{\eta}$. 
Different values of $\snr$ can be used to test CWB on regression tasks of varying hardness. 

The experimental design is defined on a grid with $n\in\{5000, 10000, 20000, 50000, 100000\}$, $p\in\{5, 10, 20, 50\}$, $p_\text{noise,rel}\in\{0.5, 1, 2, 5\}$ with $\pn = p \cdot p_\text{noise,rel}$, and $\snr\in\{0.1, 1, 10\}$. The choices of $n$, $p$, and $\pn$ are particularly relevant for memory and runtime investigations (H1) and (H4). For each combination of experimental settings, we draw 20 different target vectors for statistical replications of each scenario. When measuring the memory consumption, we only use one statistical replication of each configuration, as the memory consumption is almost identical for all repetitions of one configuration. We use \texttt{valgrind} \citep{nethercote2007valgrind} to measure the allocated memory. For memory and runtime comparisons and thus also for complexity considerations, the number of boosting iterations is set to 200. In order to assess the estimation performance of partial effects (H2) and (H3), we use the \textit{mean integrated squared error} $\text{MISE} = \frac{1}{p}\sum_{j=1}^p \int_{\min(\xj)}^{\max(\xj)}(\bj(x,\tb_k) - \bj(x, \tbh_k))^2 dx$ between the estimated base learner $\bj(\xv, \tbh_k)$ and the true effect $\bj(\xv, \tb_k)$ given by the randomly generated spline. 

For performance comparisons, we use CWB with early stopping based on a large, noise-free data set. This ensures correct early stopping and allows us to draw conclusions on CWB's estimation performance without additional uncertainty induced by finding the optimal stopping iteration. We evaluate different momentum learning rates on a uniform grid from $10^{-1}$ to $10^{-7}$. For CWB as well as for ACWB and HCWB, we set the learning rate to $\nu = 0.05$, as suggested in the literature \citep{buhlmann2007boosting}, and use $\text{df} = 5$, which provides just enough flexibility for our simulated non-linear functions. The number of boosting iterations are fixed for H1 and dynamically found for H2 and H3.

\subsubsection{Results}
\label{subsubsec:res-sim-study}

\begin{figure}[!h]
    \centering
    \includegraphics[width=0.9\textwidth]{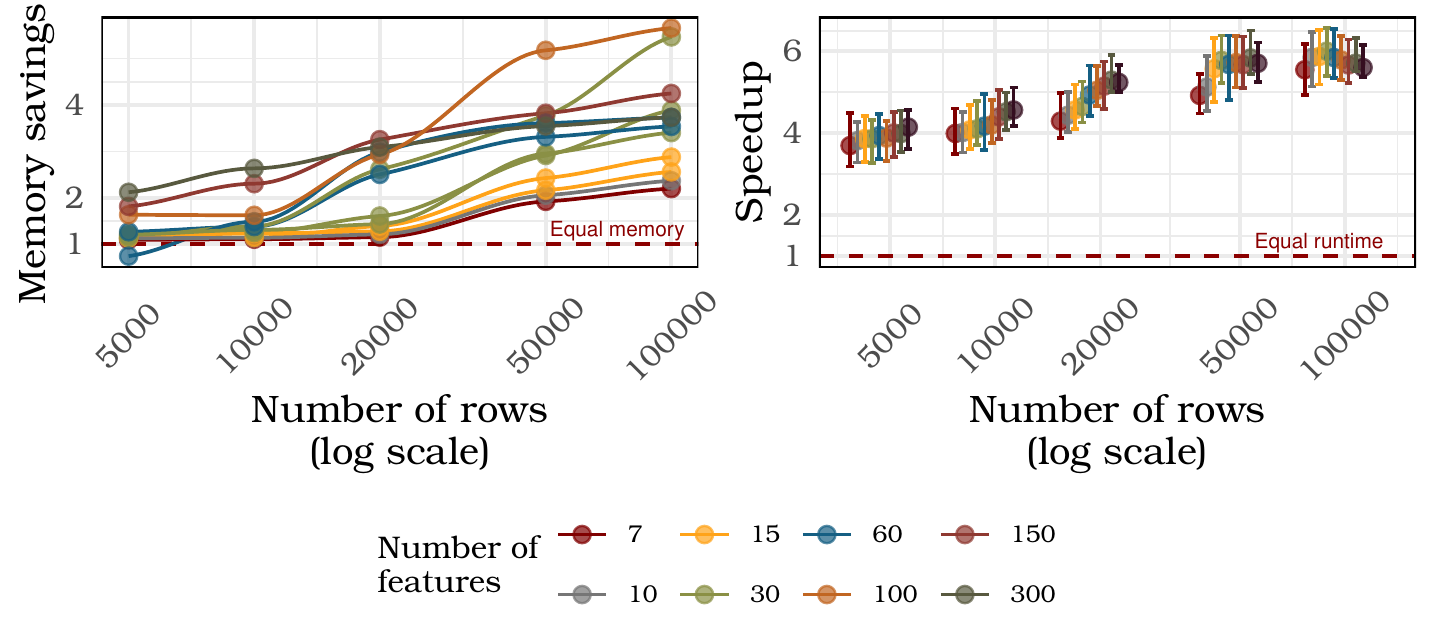}
    \caption{\footnotesize  Relative memory savings (left) and speedup as relative runtime (right) when binning is applied compared to no binning (H1). The memory savings and speedup are defined as a ratio of memory or time used by the original algorithm without binning divided by the memory consumption or runtime when binning is applied. The dashed red line indicates an equal memory consumption or runtime of both methods.}
    \label{fig:binning-performance}
\end{figure}

\textbf{H1 (memory and runtime efficiency of binning)}. Figure~\ref{fig:binning-performance} illustrates the faster runtime as well as the savings in memory consumption w.r.t. $n$ and $p$. With binning, CWB is four times faster for smaller data sets and up to six times faster for larger data sets. For smaller data sets, binning improves the memory consumption only slightly. Improvements do become large enough to be meaningful when the model is trained on larger data sets and/or more features, consuming only up to a seventh of the original memory usage. Appendix B.4.1 contains an empirical verification of the computational complexity reported in Section~\ref{subsubsec:binning-comp-complexity} using the results from H1.

\begin{figure}[h]
    \centering
    \includegraphics[width=0.9\textwidth]{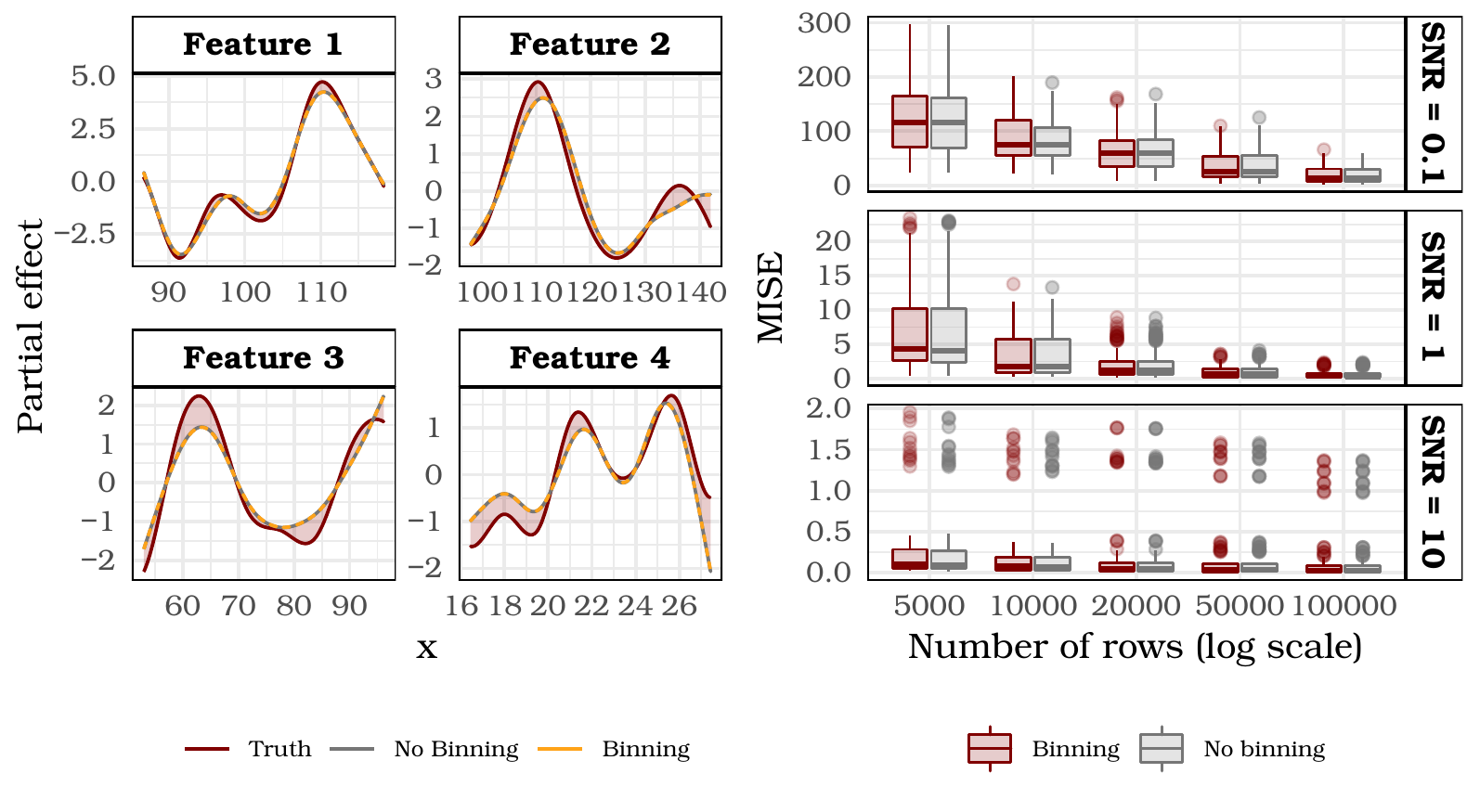}
    \caption{\footnotesize  Left: Exemplary estimated partial and true effects for $100000$ observations, $4$ features, and a SNR of $0.1$. The MISE of the curves are $49.08$ for CWB and $49.10$ for CWB with binning. Right: Comparison of the MISE when using binning vs. no binning.}
    \label{fig:binning-mise}
\end{figure}

\textbf{H2 (partial effects of binning)}. The left plot of Figure~\ref{fig:binning-mise} shows one exemplary comparison of the true effect, the estimated effect using binning, and the estimated effect without binning. The MISE is visualized by the red shaded area between the true and estimated function. As shown on the right in Figure~\ref{fig:binning-mise}, there is no notable difference between the MISE whether binning is applied or not. 

\begin{figure}[h]
    \centering
    \includegraphics[trim=0cm 0.3cm 0cm 0cm]{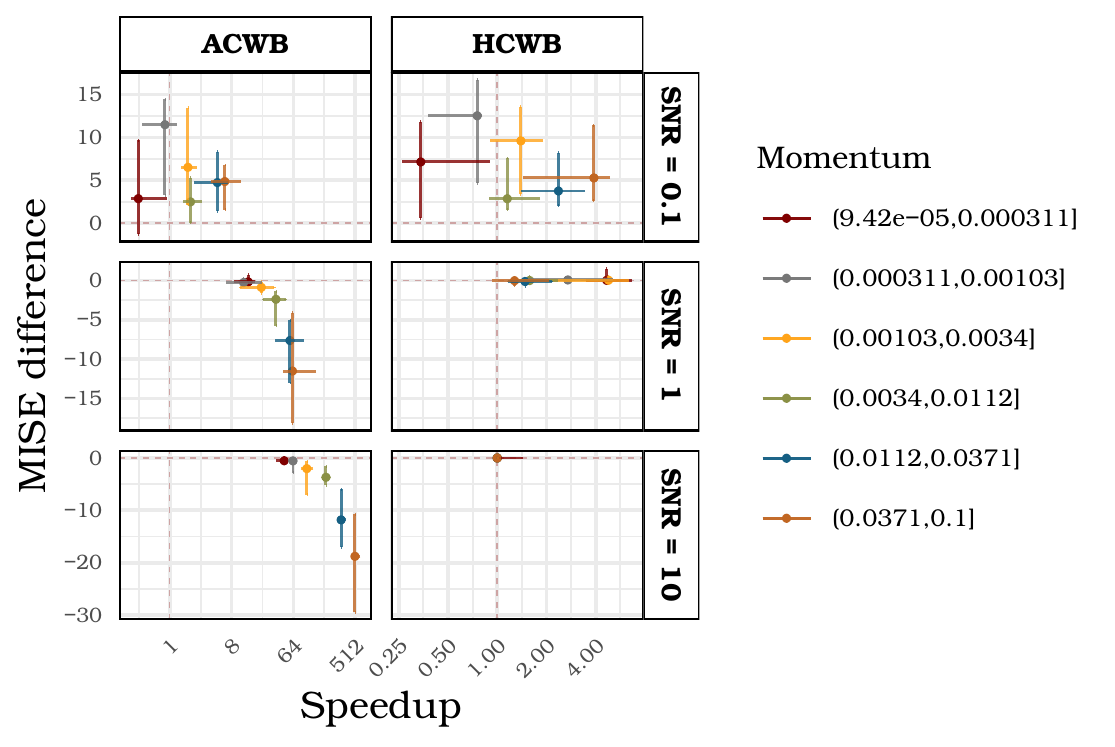}
    \caption{\footnotesize Left: MISE difference between CWB and ACWB/HCWB for partial effect estimation for different momentum values (colors) and SNR (rows). A negative difference indicates that CWB has a lower MISE and therefore a better estimation. Right: Multiplicative factor of iterations needed to train CWB compared to ACWB/HCWB, categorized by the momentum parameter (colors) and SNR (facets). A value of 10 indicates that ACWB/HCWB requires 10 times fewer iterations than CWB until convergence.}
    \label{fig:acwb-mise-iters}
\end{figure}

\textbf{H3 (iterations and partial effects of accelerated methods)}. Figure~\ref{fig:acwb-mise-iters} (left) shows the difference of MISE values of CWB compared to HCWB and ACWB. As hypothesized, partial effect estimation of ACWB is inferior to CWB due to its acceleration and potentially stopping too early or overshooting the optimal solution. This difference is negligible for small momentum values, but not for higher values. It is worth noting that the SNR in this cases is rather large, potentially undermining the effect of residual-correcting base learners of ACWB. Based on the given results, we recommend a default momentum value of $\gamma = 0.0034$ for ACWB, yielding small MISE differences while simultaneously maximizing the speedup. In contrast, HCWB performs as strongly as CWB or even better for settings with more noise (smaller SNR). The right pane of Figure~\ref{fig:acwb-mise-iters} shows the relative ratio of early stopping iterations between CBW and HCWB or ACWB. In almost all cases, the accelerated versions require fewer iterations than CWB. This is especially striking for higher momentum values, and hence we suggest a default of $\gamma = 0.037$ for HCWB. In settings with moderate or large SNR, HCWB requires as many iterations as CWB.

\textbf{H4 (scaling behavior of ACWB/HCWB)}. Next, we empirically investigate runtimes to scrutinize our derivations in Section~\ref{subsubsec:binning-comp-complexity} and~\ref{subsec:complexity-acwb}. To this end, we use the available runtimes, fit a model for each complexity statement to the given data, and compare the estimates to the theoretically derived factors in our complexity analyses. The full results are given in Appendix~B.4. In summary, for both binning and ACWB, fitted models yield an (almost) perfect fit with an $R^2$ of $1$ and $0.975$, respectively, thus underpinning our complexity estimates with only smaller deviations from theoretical numbers. In particular, this confirms the presumed speedup when using binning and the efficiency of ACWB, scaling linearly in $n$ and $K$. 

\subsection{Benchmark on Real-World Data}\label{subsec:bm-real-data}

\subsubsection{Setup}

\textbf{Algorithms:} We compare six CWB variants with XGBoost~\citep{xgboost} and EBM~\citep{nori2019interpretml}. XGBoost represents an efficient and well-performing state-of-the-art implementation of gradient boosting with tree base learners~\citep{Friedman2001}. EBM is used to compare against a recent and interpretable boosting method. Like CWB, EBM is based on additive models with additional pairwise interactions, but instead uses a different fitting technique based on a round robin selection of base learners. Although the model can be interpreted by looking at partial effects, some other key features of CWB such as unbiased feature selection cannot be directly transferred to EBM.

\textbf{Benchmark settings:} For performance comparisons in EQ2 and EQ3, we use the area under the ROC curve (AUC) based on a 5-fold cross validation (CV). Additionally, we employ a nested resampling to ensure unbiased performance estimation~\citep{bischl2012resampling} for EQ3. In the inner loop, models are tuned via Hyperband~\citep{li2017hyperband}. We use the number of boosting iterations as a budget parameter for Hyperband. We start at 39 iterations and doubled the iterations until 5000 iterations are reached. This results in 314 different HP configurations for each learner. Each of these HP configuration is evaluated using a 3-fold CV (the inner CV loop). Table~\ref{tab:bm-learner} lists the used algorithms, the corresponding software packages, and HP spaces from which each HP configuration is sampled. 

\textbf{Used software:} All experiments are executed using \texttt{R}~\citep{rmanual}. The packages used for the benchmark are \texttt{mlr3}~\citep{mlr3} as a machine learning framework with extensions, including \texttt{mlr3tuning}~\citep{mlr3tuning} for tuning, \texttt{mlr3pipelines} \citep{mlr3pipelines} for building pre-processing pipelines such as imputation or feature encoding, and \texttt{mlr3hyperband}~\citep{mlr3hyperband}. The package \texttt{interpret}, which implements EBM, was used by calling the Python implementation~\citep{nori2019interpretml} using \texttt{reticulate}~\citep{reticulate} to run EBM with its full functionality. The HP space of XGBoost is defined as the \enquote{simple set} suggested in \citet{autoxgboost}.

\spacingset{1.1}

\begin{table}[h]
    \centering
    {\scriptsize
    \begin{tabular}{|r|c|l|c|}
        \hline
        \textbf{Algorithm} & 
        \textbf{Software} & 
        \textbf{Hyperparameter space}\\
        \hline\hline
        \makecell[r]{CWB/ACWB/HCWB (nb) \\ CWB/ACWB/HCWB (b)}
            & \texttt{compboost} 
            & \makecell[l]{
                $\texttt{df} \in [2, 10]$ \\ 
                $\texttt{df\_cat} \in [2,10]$ \\ 
                $\texttt{learning\_rate} \in [0.001,0.5]$}\\
        \hline
        XGBoost 
            & \texttt{xgboost}~\citep{xgboost} 
            & \makecell[l]{
                $\texttt{eta} \in [0.001, 0.5]$ \\ 
                $\texttt{max\_depth} \in \{1, \dots, 20\}$ \\ 
                $\texttt{colsample\_bytree} \in [0.5, 1]$ \\
                $\texttt{colsample\_bylevel} \in [0.5, 1]$ \\ 
                $\texttt{subsample} \in [0.3, 1]$ \\ 
                $\texttt{lambda} \in \{2^\lambda\ |\ \lambda \in [-10,10]\}$ \\ 
                $\texttt{alpha} \in \{2^\alpha\ |\ \alpha \in [-10,10]\}$}\\
        \hline
        EBM 
            & \texttt{interpret}~\citep{nori2019interpretml} 
            & \makecell[l]{
                $\texttt{learning\_rate} \in [0.001, 0.5\}$} \\ 
        \hline
    \end{tabular}
    } 
    \caption{\footnotesize Algorithm name, software package, HP space, and number of outer evaluations of all modeling techniques compared in the benchmark.}
    \label{tab:bm-learner} 
\end{table}
    
\spacingset{1.5}

\subsubsection{Results}

\textbf{EQ1 (implementation comparison with the state-of-the-art CWB implementation \texttt{mboost})}. A comparison of our vanilla CWB implementation \texttt{compboost} with the state-of-the-art implementation \texttt{mboost} already reveals a speedup of 2 to 4 using \texttt{compboost}. When additionally using acceleration methods and binning, an increase of the speedup up to a factor of 30 for MiniBooNE and Albert can be achieved. As CWB (nb) in \texttt{compboost} and \texttt{mboost} implement the same algorithm, they are equivalent in their predictive performance. Full details are given in Appendix~B.6.

\begin{figure}[h]
    \centering
    \includegraphics[width=\textwidth]{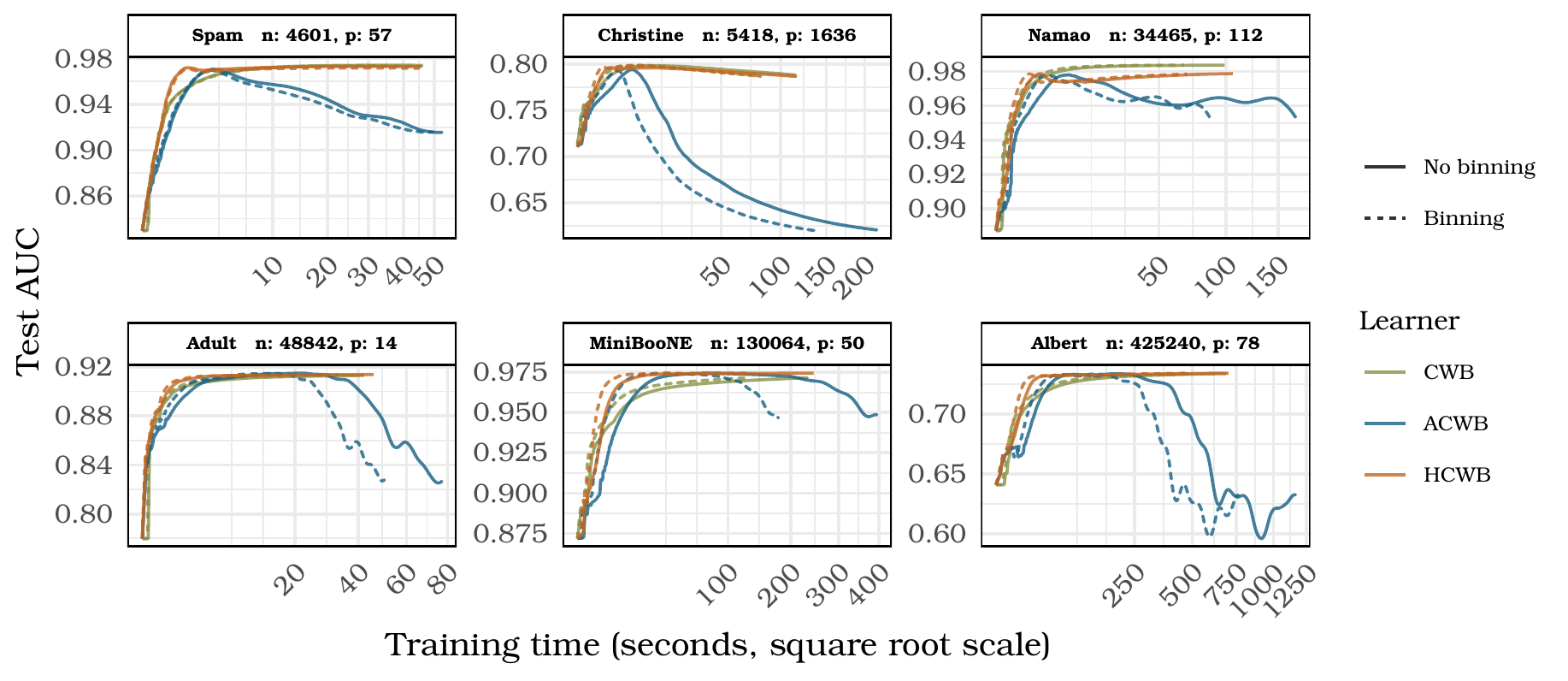}
    \caption{\footnotesize Test AUC traces of all CWB variants over 5000 boosting iterations without early stopping. Each trace is calculated as an average over the 5 runs in the 5-fold CV.}
    \label{fig:EQ2-1}
\end{figure}

\textbf{EQ2 (algorithmic comparison of accelerated methods)}. As shown in Figure~\ref{fig:EQ2-1}, HCWB (orange) learns faster than CWB (green) due to the acceleration and higher momentum. Performance improvements of ACWB (blue) take longer but surpass CWB on four out of the six data sets. As expected, the AUC of ACWB starts to decrease after the optimal number of boosting iterations is reached, while HCWB corrects this overly aggressive learning behavior by switching to CWB. The runtime of ACWB is about twice as high as for CWB due to the second error-correcting base learner fitted in each iteration (Algorithm~\ref{algo:acc-comp-boosting} line~\ref{algo:acc-comp-boosting:bl2}). Traces for binning look similar to no binning but do exhibit shorter training times, which underpins the effectiveness of binning.

\begin{figure}[h]
    \centering
    \includegraphics[width=\textwidth]{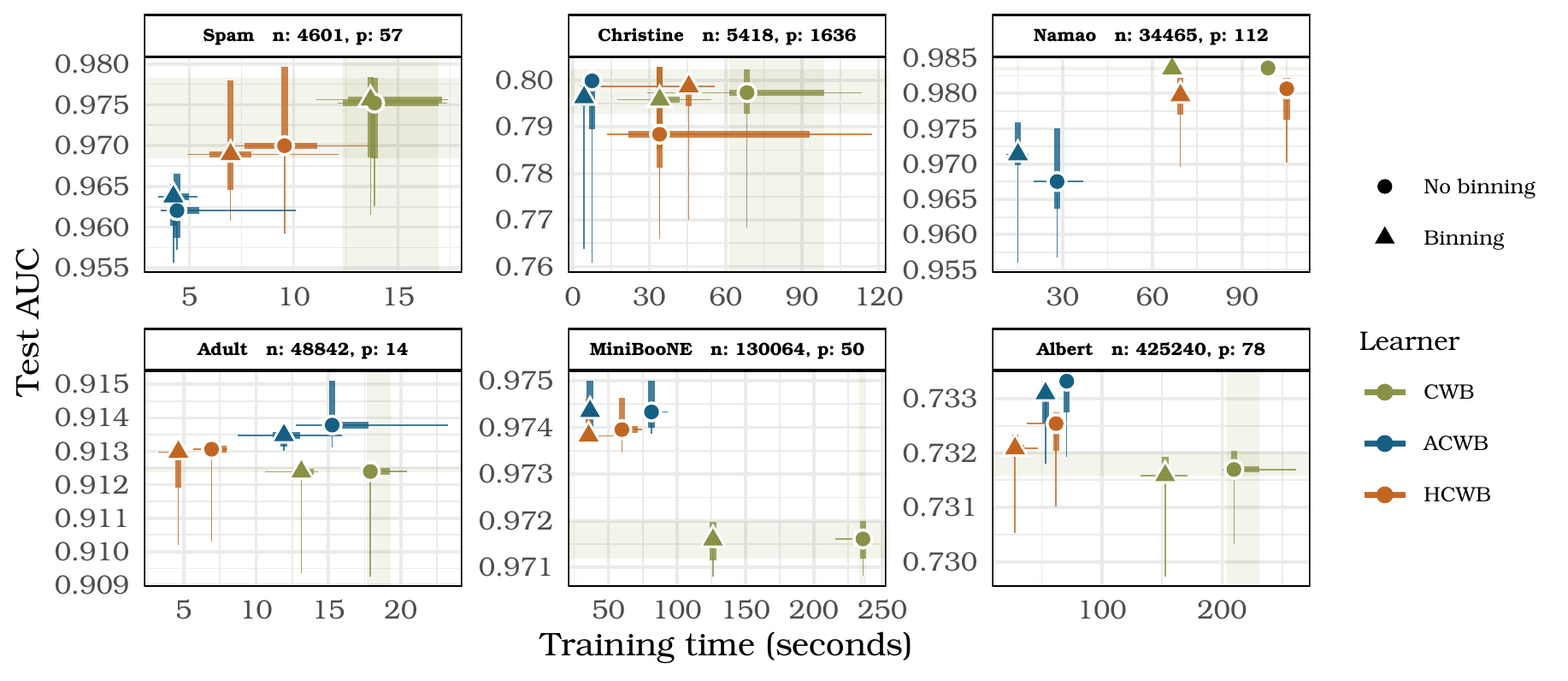}
    \caption{\footnotesize Scatter plot of average AUC values and training times for all CWB variants (color / symbol). Horizontal and vertical boxes indicate the 25- and 75\%-quantile, and colored lines indicate the possible range of values. Light green areas additionally highlight the 25- and 75\%-quantiles of CWB (nb) to facilitate easy comparison of other CWB variants with this baseline.}
    \label{fig:EQ2-2}
\end{figure}

Figure~\ref{fig:EQ2-2} additionally shows test AUC and runtimes of all CWB variants when using early stopping based on a validation data set (defined as 30\% of the training data). To check if performance changes between different models are significant, we use the resulting AUC values and compute a beta regression with learners as covariates and the AUC as a response variable (see Appendix B.8). Both ACWB (p-value $= 0.4122$) and HCWB (p-value $= 0.6927$) do not yield a significantly smaller AUC value. Furthermore, results show that binning does also not have a significant effect on the performance (p-value $= 0.9594$). At the same time, binning improves the runtime by an average speedup of 1.5 for all three CWB variants (cf. Appendix~B.7 Table~3). ACWB and HCWB even yield further improvements with an average speedup of 3.8 and 2.38, respectively.

\begin{figure}[h]
    \centering
    \includegraphics[width=\textwidth]{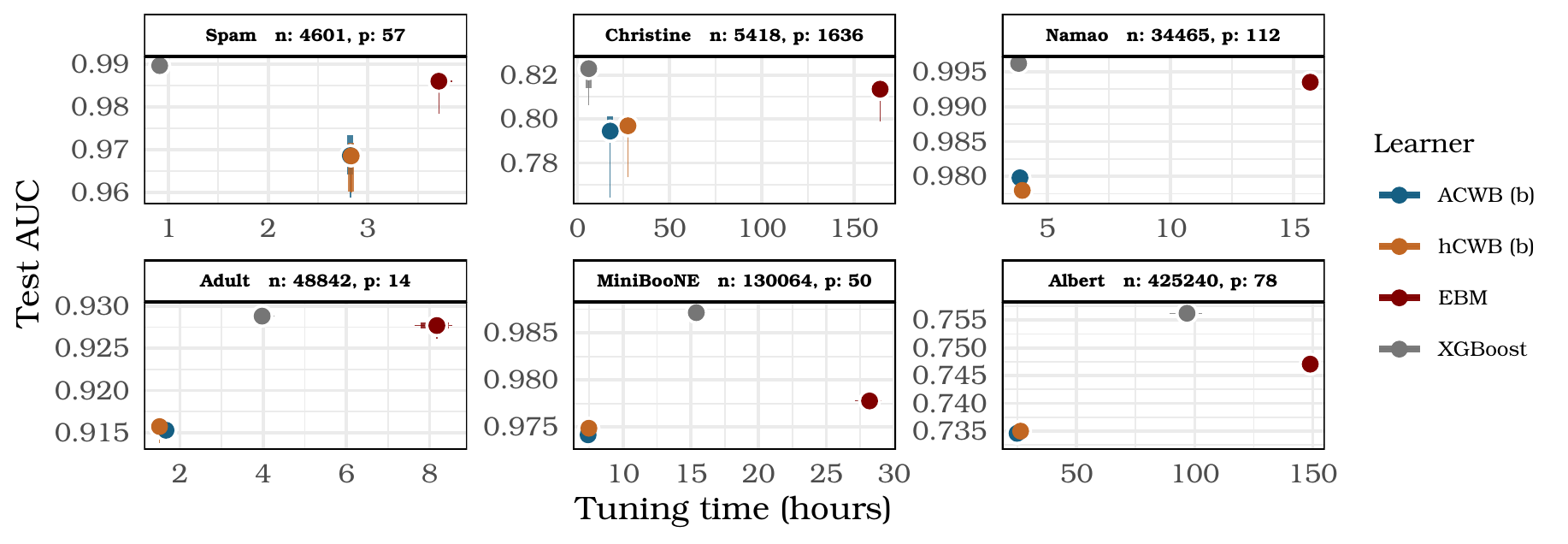}
    \caption{\footnotesize AUC and training time of ACWB (b), HCWB (b), XGBoost, and EBM.}
    \label{fig:EQ3}
\end{figure}

\textbf{EQ3 (comparison with state-of-the-art algorithms)}. Figure~\ref{fig:EQ3} shows that state-of-the-art algorithms benefit from their more complex model structure by also considering complex feature interactions (EBM allows for interactions by design, while XGBoost uses tree base learners, which induce more complex interactions with larger tree depth). The improvement in AUC of these methods compared to CWB is only practically relevant for the data set Christine, which shows an AUC increase of 3.42\% for XGBoost. The AUC improvement for all other tasks is (notably) smaller than 3\%, even though our approach uses a fully interpretable model. In terms of runtime, ACWB (b) and HCWB (b) outperform XGBoost and EBM on most data sets. On the data sets Spam and Christine, XGBoost is faster than our algorithms, which we attribute to their small sample size. In general, ACWB (b) and HCWB (b) are 4.62 times faster than EBM and 1.66 times faster than XGBoost. 

\section{Conclusion}

Adaptions to CWB presented in this paper can notably improve the use of computational resources, reducing runtimes by up to a factor of 6 and memory usage up to factor of 4, depending on the size of the task. Incorporating binning with equally spaced design points efficiently leverages CWB's base learner structure with one feature per base learner. Benchmark results show that binning reduces the training time without impairing the predictive performance. The proposed accelerated CWB algorithm (ACWB) is furthermore a natural extension of CWB and provides a faster training procedure at the expense of a potential deterioration of predictive performance when not stopped properly. Our alternative hybrid solution (HCWB) accounts for this and presents no drawbacks in comparison to the standard CWB algorithm. However, HCWB does incur slightly longer runtimes compared to ACWB. In practice, HCBW yields good out-of-the-box performance, while ACWB does not require an additional validation data set and can thus be beneficial in low sample size regimes. 
Future research will investigate how the proposed framework can be used and further improved for more complex additive models structures in both the predictors as well as in the outcome, e.g., for functional regression models.


\bigskip
\begin{center}
{\large\bf ACKNOWLEDGMENT}
\end{center}

\begin{footnotesize}
This work was supported by the German Federal Ministry of Education and Research (BMBF) under Grant No. 01IS18036A and Federal Ministry for Research and Technology (BMFT) under Grant FKZ: 01ZZ1804C (DIFUTURE, MII). The authors of this work take full responsibilities for its content.
\end{footnotesize}

\bigskip
\begin{center}
{\large\bf SUPPLEMENTARY MATERIAL}
\end{center}

\begin{footnotesize}

\begin{description}

\item[Appendix:] Descriptions of possible categorical feature representations with a short comparison w.r.t. runtime and memory consumption as well as class selection properties in the presence of noise. The Appendix further contains empirical validation of the computational complexity estimates as given in Section~\ref{subsec:cwb-complexity} and~\ref{subsubsec:binning-comp-complexity}. Lastly, the appendix contains a figure showing the full benchmark, as well as a generalized linear model to statistically investigate the effect of our adaptions on the predictive performance. (PDF file)

\item[Source code of \texttt{compboost}:] \url{github.com/schalkdaniel/compboost} (Commit tag of the snapshot used in this paper: \texttt{c68e8fb32aea862750991260d243cdca1d3ebd0e})

\item[Benchmark source code:] \url{https://github.com/schalkdaniel/cacb-paper-bmr}

\item[Benchmark Docker:] Docker image with pre-installed packages to run the benchmark and access results for manual inspection: \url{hub.docker.com/repository/docker/schalkdaniel/cacb-paper-bmr}

\end{description}
\end{footnotesize}


\bibliographystyle{chicago}
\setlength{\bibsep}{0pt plus 0.3ex}
\begin{footnotesize}
\bibliography{references}
\end{footnotesize}

\end{document}


\def\spacingset#1{\renewcommand{\baselinestretch}%
{#1}\small\normalsize} \spacingset{1}

\date{}

\if0\blind
{
  \title{\bf Supplementary Material for Accelerated Component-wise Gradient Boosting using Efficient Data Representation and Momentum-based Optimization}
  \author{Daniel Schalk, 
    Bernd Bischl 
    and 
    David Rügamer \\
    Department of Statistics, LMU Munich}
  \maketitle
} \fi

\begin{appendix}

\section{Further Theoretical Details}

\subsection{Vanilla CWB Algorithm}

\begin{algorithm}
\caption{Original CWB algorithm given the input and output.}\label{algo:cboost-basic} 
\vspace{0.15cm}
\hspace*{\algorithmicindent} \textbf{Input} Train data $\D$, learning rate $\nu$, number of boosting iterations $M$, loss function\\
\hspace*{\algorithmicindent} \phantom{\textbf{Input} }$L$, set of base learner $\mathcal{B}$\\
\hspace*{\algorithmicindent} \textbf{Output} Prediction model $\hat{f}^{[M]}$ defined by fitted parameters $\tbih{1}, \ldots, \tbih{M}$\vspace{0.15cm}
\hrule

\begin{algorithmic}[1]
\Procedure{CWB}{$\D,\nu,L,\mathcal{B}$}
    \State Initialize: $\fh^{[0]}(\xv) = \argmin_{c\in\R}\riske(c)$
    \For{$m \in \{1, \dots, M\}$}
        \State $\rmi = -\left.\fp{\Lxyi}{f(\xi)}\right|_{f = \fmdh},\ \ \forall i \in \{1, \dots, n\}$ \label{algo:cboost-start} 
        \For{$k \in \{1, \dots, K\}$}\label{algo:cboost-findbl-start}
            \State $\tbmh = \argmin_{\tb\in\R^{\bdj}} \sum_{i=1}^n\left(\rmi - \bj(\xi,\tb)\right)^2$\label{algo-cboost-basic-blearner-fit} 
            \State $\text{SSE}_k = \sum_{i=1}^n(\rmi - \bj(\xi, \tbmh))^2$ 
        \EndFor\label{algo:cboost-findbl-end}
        \State $k^{[m]} = \argmin_{k\in\{1, \dots, K\}} \text{SSE}_k$ 
        \State $\fmh(\xv) = \fmdh(\xv) + \nu \bjm (\xv,\tbmh)$ \label{algo:cboost-end} 
    \EndFor
    \State \textbf{return} $\fh = \fh^{[M]}$
\EndProcedure
\end{algorithmic}
\end{algorithm}

\subsection{Unbiased Feature Selection}\label{app:unbiased-fs}

CWB also can incorporate penalties into each base learner. A bigger penalization could lead to inflexible base learners having a disadvantage in being able to fit the pseudo residuals compared to other base learners. This leads to a preference of selecting flexible base learner more often than inflexible ones. Hence, it is desirable to allow for a fair selection between more and less flexible base learners \citep{hofner2011framework}. Let $\bj$ be a penalized regression model base learner with design matrix $\design_k$, penalty matrix $\bm{D}_k\in\R^{\bdj\times \bdj}$, an optional diagonal weight matrix $\bm{W}_k\in\R^{n\times n}$ as well as an optional smoothing parameter $\lambda\in\R$. Additionally, we assume that $\bm{W}_k$ and $\bm{D}_k$ are symmetric and can be decomposed into $\bm{W}_k = (\bm{W}_k^{1/2})^\tran\bm{W}_k^{1/2}$ and $\bm{D}_k = (\bm{D}_k^{1/2})^\tran \bm{D}_k^{1/2}$. The reason for that decomposition becomes clear when solving the objective to fit penalized base learner to the pseudo residuals:
\begin{equation}\label{eq:pen-regr}
    \tbmh = \argmin_{\tb\in\R^\bdj}\left(\left\|\rmm - \bm{W}_k^{1/2}\design_k\tb\right\|_2^2 + \lambda\left\|\bm{D}_k^{1/2}\tb\right\|_2^2\right).
\end{equation}
Here, $\|\cdot\|$ denotes the Euclidean norm. The fitted pseudo residuals $\rmh_k\in\R^n$ are given by 
\begin{equation}
\rmh_k = \design_k(\design_k^\tran\bm{W}_k\design_k + \lambda \bm{D}_k)^{-1}\design_k^\tran \bm{W}_k \rmm = \bm{H}_k(\lambda) \rmm \label{modelfit}
\end{equation}
where the decompositions $\bm{D}_k^{1/2}$ and $\bm{W}_k^{1/2}$ are aggregated to $\bm{D}_k$ and $\bm{W}_k$.\\

Common examples for $\bm{D}_k^{1/2}$ is the identity matrix $\bm{D}_k^{1/2} = I_n \in\R^{\bdj\times \bdj}$ used for the ridge regression or the second order difference matrix $\bm{D}_k^{1/2}(i) = (0, \ldots, 0, 1, -2, 1, 0, \ldots, 0),\ \bm{D}_k^{1/2} \in\R^{\bdj-2 \times \bdj}$, with $i-1$ zeros before $1$, $-2$, and $1$, used for P-splines. Common examples for $\bm{W}_k$ is using observational weights on the diagonal $\bm{W}_k = \diag\left(w_k^{(1)}, \ldots, w_k^{(n)}\right)$ and $\bm{W}_k^{1/2} = \diag\left((\sqrt{w_k^{(1)}}, \ldots, \sqrt{w_k^{(n)}}\right)$ or correcting for heteroscedasticity with $\bm{W}_k = \diag(1/\sigma_1^2, \ldots, 1/\sigma_n^2)$ and $\bm{W}_k^{1/2} = \diag(1/\sigma_1, \ldots, 1/\sigma_n)$ correcting for an observational based variance.\\ 

In order to ensure an unbiased selection of base learners of different complexity, the degrees of freedom of the $j$-th base learner fit $\rmh_k$ are used and set equally for each base learner. Two common versions of degrees of freedom are defined via the trace of the {hat matrix} $\bm{H}_k(\lambda)$ \citep[see, e.g.,][]{buja1989linear}:
\begin{align}
    \mbox{df}_1 &= \trace(\bm{H}_k(\lambda)) \label{df1}, \\
    \mbox{df}_2 &= \trace(2\bm{H}_k(\lambda) - \bm{H}_k(\lambda)\bm{H}_k(\lambda)).  \label{df2}
\end{align}
The one-to-one relationship allows to define equal flexibility of each base learner in practice by defining the respective penalization such that all base learners have the same degrees of freedom. In general, there is no analytic solution for solving \eqref{df1} or \eqref{df2} for $\lambda$ and the problem must be solved numerically. A naive uniroot search is in many cases too expensive since $\lambda$ is part of the inverse matrix $(\design_k^\tran\bm{W}_k\design_k + \lambda \bm{D}_k)^{-1}$ in \eqref{modelfit} which would then have to be calculated for every $\lambda$ value anew. To avoid this, the Demmler-Reinsch-Orthogonalization \citep[DRO;][]{ruppert2002selecting} is used.

\subsection{Categorical Features}\label{app:cat-features}

Categorical features $\xj$ consisting of $c_j$ classes $\{1, \dots, c_j\}$. These classes do not necessarily need to be encoded as integer values, but we keep this formalization due to simplicity and unified notation. For example, a categorical feature for three different prescribed medications consists of three classes  $1 = \textit{Medication 1}$, $2 = \textit{Medication 2}$, and $3 = \textit{Medication 3}$. \\

The basic idea of efficient categorical feature representations is to make use of the number of classes $c_j$ which is much smaller than the number of observations $n$. Working on the number of classes allows us to fit base learner of categorical features faster and with less memory usage than operating on the instances. We describe two common representation for CWB, highlight their advantages, and how to interpret the estimated base learner. \\

\subsubsection{Ridge Representation}

One possibility to include categorical features into a base learner is to make use of techniques applied in linear models. The usual way to encode categorical features is to use a dummy-encoding represented by a binary model matrix $\bm{X}_j\in\mathbb{R}^{n\times c_j}$ with elements $\bm{Z}_j(i,k) = \mathds{1}_{\{\xij = k\}}$, $k \in \{1, \dots, c_j\}$. To obtain the regression coefficients of a base learner with categorical feature $\xj$, we use ridge regression~\citep{hoerl1970ridge} and solve
\begin{equation}
    \tbmh = \argmin_{\tb\in\R^{c_j}} \|\rmm - \bm{X}_j\tb\|_2^2 + \lambda\|\tb\|_2^2
    \label{eq:penLS}
\end{equation}
where $\lambda\in\R_+$ is a penalty which additionally shrinks the coefficients. Equation \eqref{eq:penLS} can be solved analytically with solution $\tbmh = (\bm{Z}_j^\tran\bm{Z}_j + \lambda\bm{I}_{c_j})^{-1}\bm{Z}_j^\tran \rmm$ where $\bm{I}_{c_j}\in\R^{c_j \times c_j}$ is the identity matrix. Advantages of this representation are:

\begin{myIndent}
\textbf{Efficient fitting procedure} that analytically calculates the inverse of $\bm{X}_j^\tran\bm{X}_j + \lambda\bm{I}_{c_j}$ by making use of the diagonal structure. The inverse is calculated as $\diag((n_{j,1} + \lambda)^{-1}, \dots, (n_{j,c_j} + \lambda)^{-1})$ where the elements $n_{j,k} = \sum_{i=1}^n \mathds{1}_{\{\xij = k\}}$ correspond to the cardinality of class $k$ of the $j$-th feature. Instead of storing the binary matrix $\bm{X}_j$ which requires at least $2n$ integer values for a sparse data format, we store $c_j$ double values representing the diagonal. Additionally, we do not have to calculate the inverse in each iteration. Furthermore, the matrix multiplication $\bm{u} = \bm{X}_j\bm{r}^{[m]}$ can be efficiently calculated by group means $u_k = \sum_{i=1}^n \mathds{1}_{\{x_{i,j} = k\}}\rmi$, $k \in\{1, \dots, c_j\}$.
\end{myIndent}

\begin{myIndent}
\textbf{Explicit calculation of the degrees of freedom} to conduct an unbiased feature selection between base learners even if the base learners are not from the same type (e.g. comparison of spline and ridge base learner). The calculation is done by making use of the diagonal structure. As mentioned in Section~\ref{app:unbiased-fs} we have to use the DRO which in turn requires a \textit{singular value decomposition} (SVD) to calculate the degrees of freedom. Because of the diagonal structure, it is not necessary to calculate the SVD explicitly. The degrees of freedom are directly given as function of the number of observations:
\begin{align}
\mbox{df}_1 &= \sum_{k=1}^{c_j} \frac{n_{j,k}}{(n_{j,k}+\lambda)} \\
\mbox{df}_2 &= \sum_{k=1}^{c_j} \frac{n_{j,k} (n_{j,k} + 2\lambda)}{(n_{j,k}+\lambda)^2}  \label{df2sol}.
\end{align}
\end{myIndent}

\begin{myIndent}
\textbf{Number of base learner to loop over} is independent from the number of classes. Using the ridge representation constructs one base learner per categorical feature. Therefore, the number of base learners and the complexity of the model does not increase with increasing numbers of classes $c_j$ as it is for the binary representation in the next section.  
\end{myIndent}

\subsubsection{Sparse binary Representation}

Instead of having one base learner per feature, another representation can be applied on the class level. Therefore, each class $k \in \{1, \dots, c_j\}$ of the categorical feature $\xj$ is a new base learner containing a binary vector just for that class. The model matrix $\bm{X}_j$ then is of dimension $n\times 1$ and contains ones at the respective entries and otherwise zeros. The parameter estimate is again calculated using the method of least squares 
\begin{equation}
    \hat{\theta}_{j,k}^{[m]} = (\bm{X}_j^\tran \bm{X}_j)^{-1}\bm{X}_j^\tran \rmm = n_{j,k}^{-1}\sum\limits_{i=1}^n \mathds{1}_{\{x_{i,j} = g\}} \rmi.
\end{equation} 
Following this definition, the coefficient $\hat{\theta}_{j,k}$ is a scalar representing the $k$-th class mean of feature $j$. Advantages of this representation are as follows:

\begin{myIndent}
\textbf{Automated class selection} within the categorical feature. Compared to updating mechanisms where all parameter are updated at once (i.e., as for the ridge representation) the binary representation selects just one class per iteration. This yields an automatic class selection induced by the fitting process. Further information about importance of classes can then be derived from the trace on how the classes are selected. This is also helpful when one of the classes does not contain information and therefore should not be selected.
\end{myIndent}

\begin{myIndent}
\textbf{A sparser model is obtained} due to individual class selections within one categorical feature.
\end{myIndent}

\section{Further Experimental Findings and Details}

\subsection{Simulation Study - Categorical Features}

We do not compare the two categorical encodings explained in this paper since there is no state-of-the-art encoding to compare with. Instead, we want to highlight their computational properties and also how to interpret the estimated base learner with the respective encoding.\\

\begin{figure}[h]
    \centering
    \includegraphics[width=0.45\textwidth]{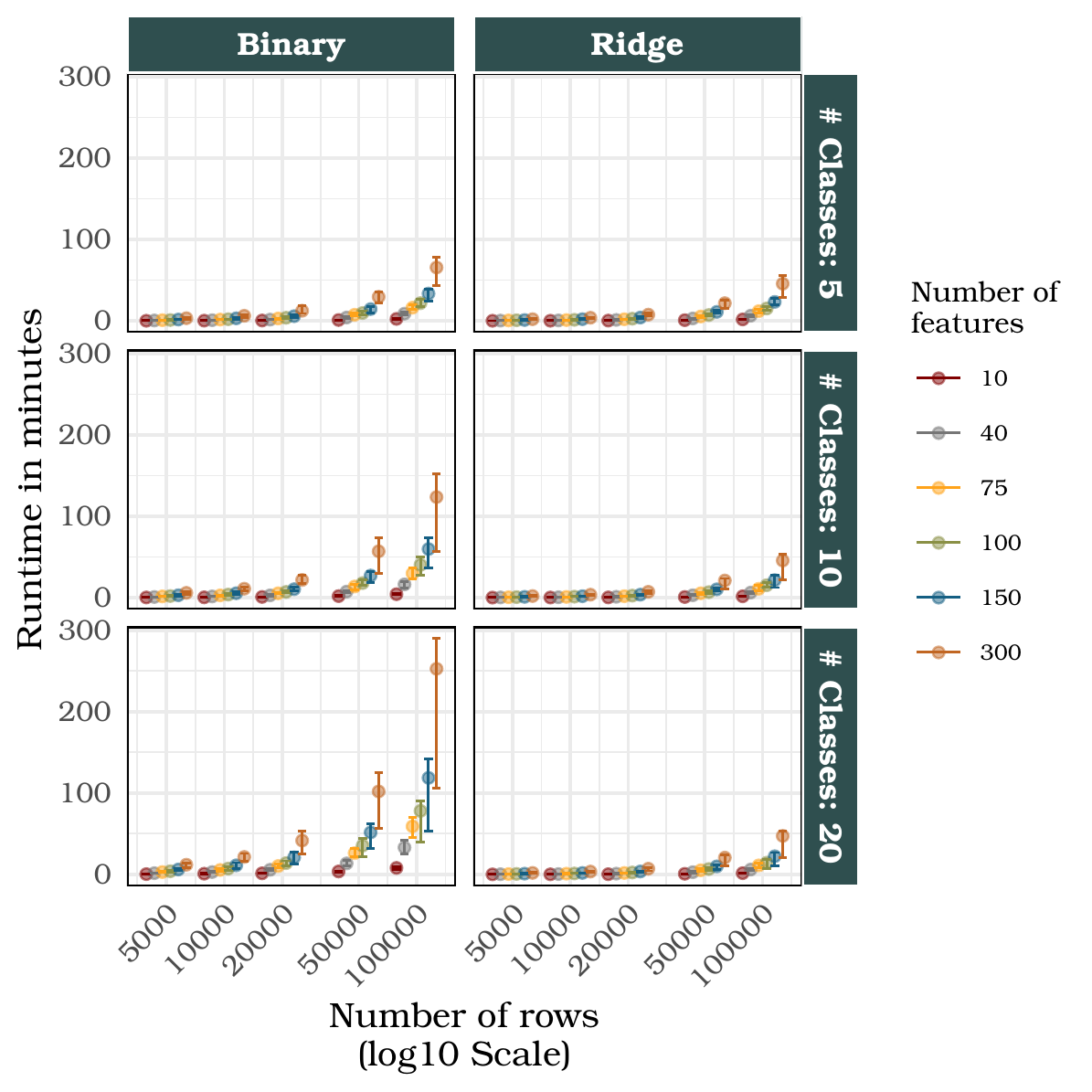}
    \includegraphics[width=0.45\textwidth]{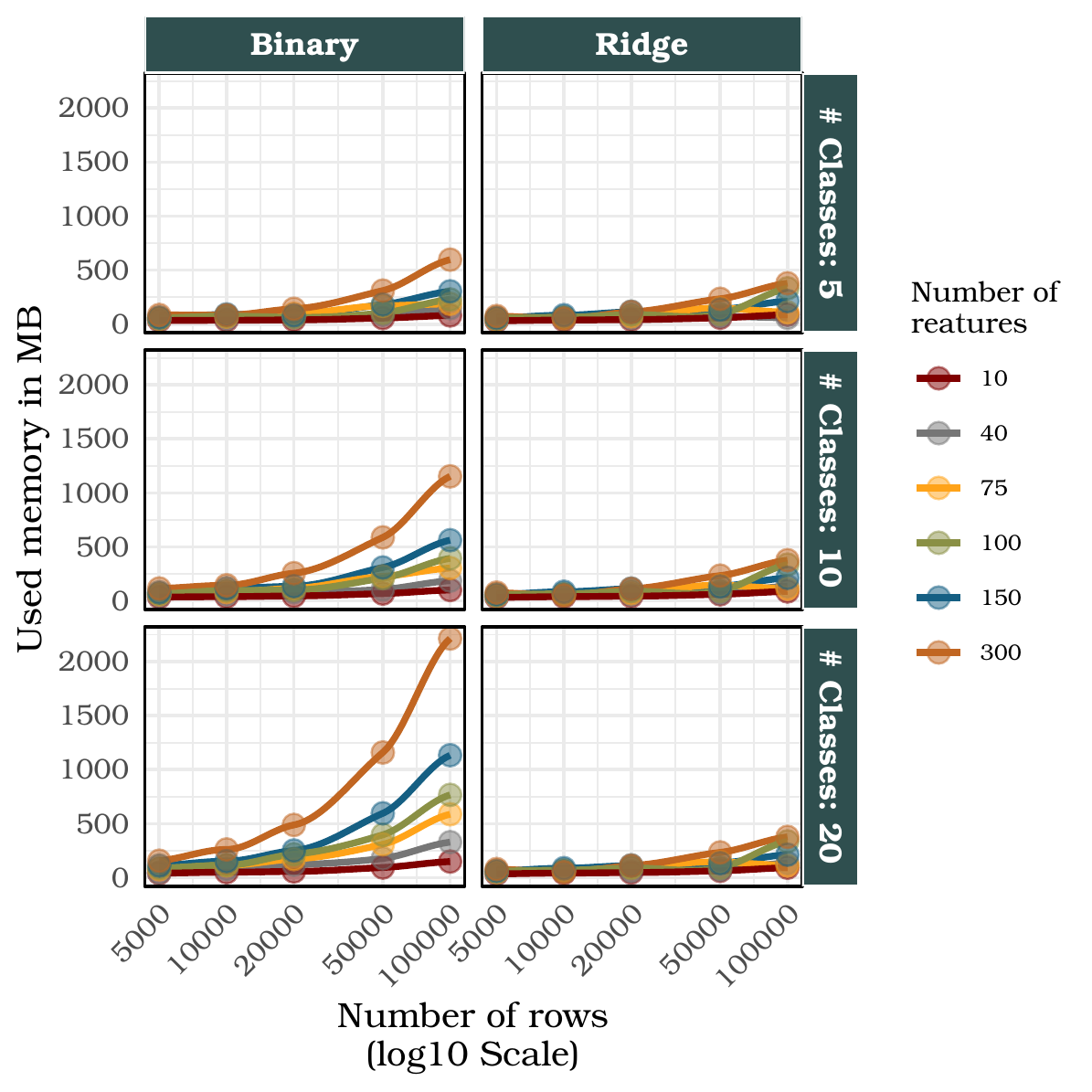}
    \caption{Runtime in seconds (left). Memory consumption in megabyte (right).}
    \label{fig:cat-memory-runtime}
\end{figure}

As shown in Figure~\ref{fig:cat-memory-runtime}, the ridge encoding is much faster in terms of runtime and memory. The effect can be explained by again looking at the structure of the base learners. Using the binary encoding requires to fit as many base learners as classes in the feature. The high memory consumption when using binary base learner can be explained by looking at the metadata the base learner stores. Each binary base learner holds a vector of indexes to calculate the group mean in each iteration. The reason for that is to not loop over all $n$ feature values in each iteration for each binary base learner but just over the subset of feature values corresponding to the specific class.  In contrast, the ridge base learner stores one vector with the number of observations per class and uses the original feature value to calculate the group means. \\

\begin{figure}[h]
    \centering
    \includegraphics[width=0.6\textwidth]{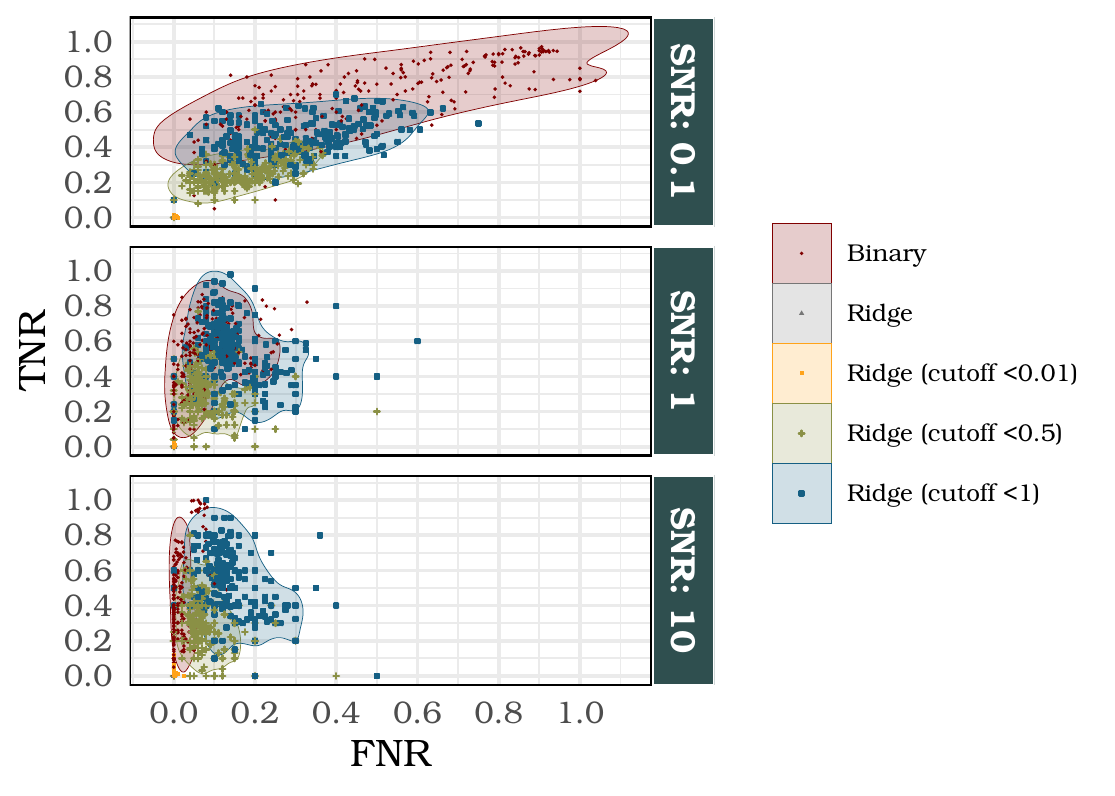}
    \caption{Fraction of being able to filter non-informative classes (TNR; class was not selected and true parameter is zero) and wrongly filtering informative classes (FNR; class was not selected but true parameter is not zero). One point in the figure corresponds to the median of the 20 replications of one configuration. The optimal point would be in the left upper corner at $(0,1)$, meaning that all non-informative classes were filtered and all informative classes were  selecting all classes with a signal. The contour lines are the two dimensional empirical 0.95 quantiles.}
    \label{fig:cat-selection}
\end{figure}

The binary encoding surpasses the ridge encoding in terms of filtering non-informative classes. As we can see in Figure~\ref{fig:cat-selection}, the binary encoding is able to filter non-informative classes while the ridge representation does not filter them. Note that the points for the ridge encoding are all placed at $(0,0)$ due to the simultaneous updating of all class parameter. In contrast, the binary representation updates just one class parameter per iteration. The cost for a better TNR comes with the risk of not selecting important features (higher FNR) for a higher SNR. Hence, the binary encoding has a very conservative selection process, but if a class is selected it was likely selected due to a signal present in the data. \\

\begin{figure}[ht]
    \centering
    \includegraphics[width=0.6\textwidth]{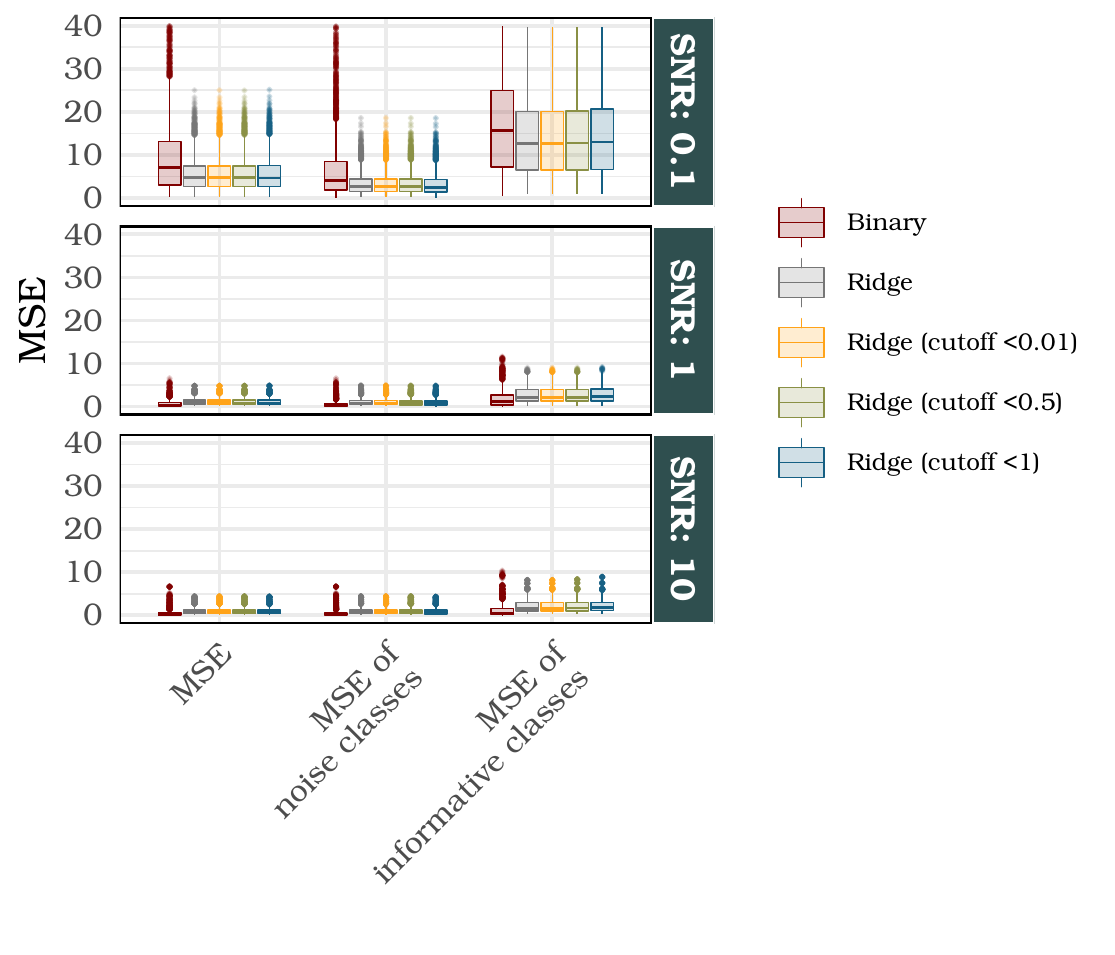}
    \caption{MSE of the estimated parameter the true ones. The MSE is shown for all parameter (MSE), parameters corresponding to noise classes, and parameter corresponding to informative classes (MSE of informative classes)}
    \label{fig:cat-ridge-parameter-mse}
\end{figure}

As shown in Figure~\ref{fig:cat-ridge-parameter-mse}, the MSE of the estimated parameter and the true ones is small for a smaller SNR but increases for more complex situations. The MSE of the binary encoding is slightly better as for the ridge representation but is slightly worse for the high SNR case. Even though the small MSE of the parameter of non-informative classes, they appear in the model and therefore increase the complexity. This is especially the case for ridge regression. A strategy to overcome this issue is to set a threshold for which the smaller parameters are set to zero. As we can see in Figure~\ref{fig:cat-ridge-parameter-mse}, this cutoff does not affect the MSE but leads to a better TNR than ridge regression (see Figure~\ref{fig:cat-selection}). The challenge now is to accordingly set this cutoff value to not cut parameters too aggressively and therefore ignore informative classes. Figures~\ref{fig:cat-selection} and~\ref{fig:cat-ridge-parameter-mse} contains this strategy for a cutoff of 0.01, 0.5, and 1. As we can see, the TNR is improves while the FNR gets worse. Looking at Figure~\ref{fig:cat-ridge-parameter-mse} in the context of Figure~\ref{fig:cat-selection}, we can explain the higher MSE for binary base learner for a SNR of 0.1 due to the higher FNR and therefore a higher MSE on the not selected classes.  \\

To summarize the results, it is much faster and more memory friendly to use the ridge representation. The drawback of the ridge representation is the risk of wrongly selecting non-informative classes. Despite the fact that we can cut off classes, the way binary encoding selects classes is more natural. If computational resources are an issue, we suggest to use the ridge representation while the binary representation gives us a sparser model and a more precise selection of classes.

\subsection{Used Real World Data Sets}

\textbf{Data sets:} We use 6 data sets from OpenML \citep{OpenML2013,OpenMLR2017} for binary classification and provide descriptive statistics in Table~\ref{tab:res-bm-datasets}. The data sets are pre-processed by imputing missing values in numerical features with the median, in categorical features with the mode, and then removing constant features from the data\footnote{Note that we keep thing simple here by applying all preprocessing to the complete data sets, which does not affect the validity of our comparison experiments. In proper applied work, such pre-processing should be embedded into cross-validation.}. 

\begin{table}[ht]
    \centering
    {\scriptsize
    \begin{tabular}{|r|c|c|c|c|c|}
        \hline
        \multirow{2}{*}{\textbf{Data set}} & 
        \multirow{2}{*}{\textbf{Data ID}} & 
        \multirow{2}{*}{\textbf{\# Samples}} & 
        \multicolumn{2}{c|}{\textbf{\# Features}}\\ 
        & & & Numeric & Categorical\\ 
        \hline\hline
        Spam & 44 & 4601 & 57 & 0\\
        \hline
        Christine & 41142 & 5418 & 1599 & 37 \\
        \hline
        Namao & 1486 & 34465 & 83 & 29 \\
        \hline
        Adult & 1590 & 48842 & 6 & 8 \\
        \hline
        MiniBooNE & 41150 & 130064 & 50 & 0\\
        \hline
        Albert & 41147 & 425240 & 26 & 52\\
        \hline
    \end{tabular}
    } 
    \caption{\footnotesize Key characteristics of the used data sets.}
    \label{tab:res-bm-datasets} 
\end{table}

\subsection{Picture of varying Degrees of Freedom and Number of Bins}

\begin{figure}[ht]
    \centering
    \includegraphics[width=\textwidth]{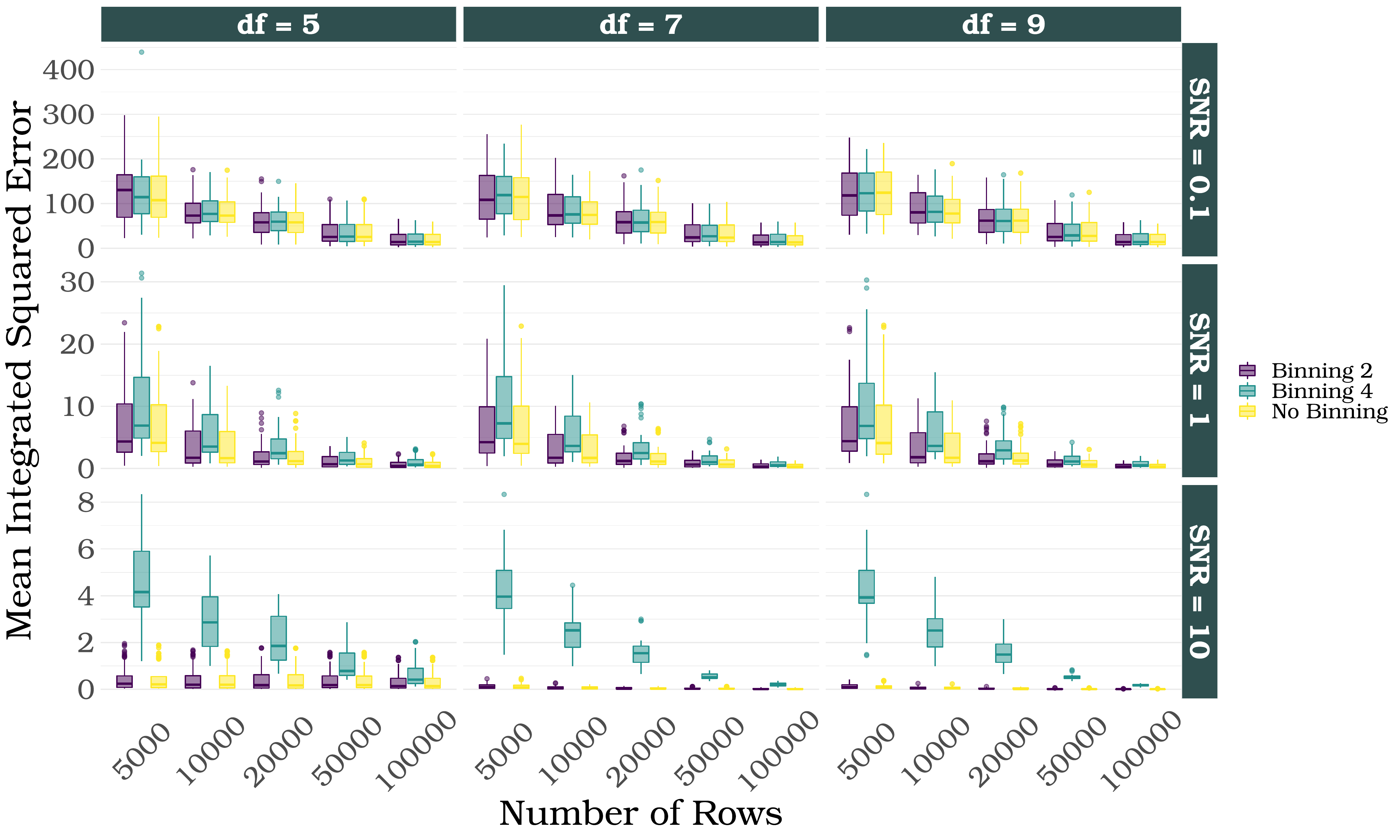}
    \caption{MISE of binning when applied with $n^\ast = n^{1/2}$ and $n^\ast = n^{1/4}$ as well as $\text{df} \in \{5, 7, 9\}$.}
    \label{fig:binning-df-nroot}
\end{figure}

To get further insights how the choice of number of bins affects the performance, we conduct the benchmark with $n^\ast = n^{1/4}$. Additionally, we choose different values for the degrees of freedom. The results are shown in Figure~\ref{fig:binning-df-nroot}. 

\subsection{Empirical Assessment of Computational Complexity}

\subsubsection{Binning}

\begin{figure}[ht]
    \centering
    \includegraphics[width=0.7\textwidth]{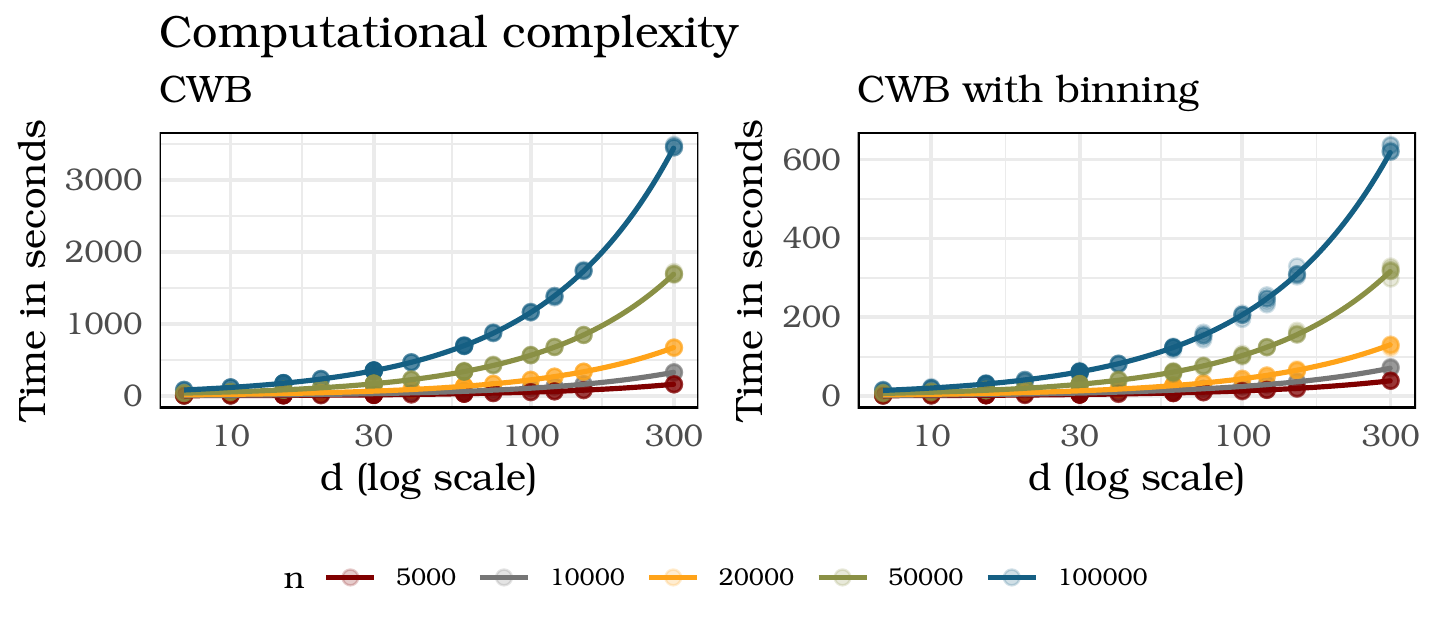}
    \caption{\footnotesize  Fitted curves based on computational complexity with time used as proxy and response variable. The number of rows and base learner are used as input.}
    \label{fig:cwb-complexity-est}
\end{figure}

Modeling the time as proxy for the computational complexity as calculated in Section 2.4 for CWB and 3.1.3 for CWB with binning confirmes our complexity estimates with $\mathcal{O}(K^{0.989}(\bd^{2.014}n^{1.030} + \bd^{2.985}) + M^{1.025}K^{1.006}(\bd^{2.040} + \bd^{0.992}n^{0.982}))$ for CWB and respectively $\mathcal{O}(K^{1.034}(\bd^{1.989}(n^\ast)^{0.994}  + n^{0.956} + \bd^{3.065}) + M^{1.020}K^{0.992}(\bd^{2.000} + \bd^{1.005} (n^\ast)^{0.998} + n^{0.993}))$ for CWB with binning. Figure~\ref{fig:cwb-complexity-est} shows the fitted curves. Additionally, a scale parameter $v = 3.989\cdot 10^7$ for CWB and $v = 1.072\cdot 10^7$ for CWB with binning was estimated to account for different scales of seconds and number of operations. The R-square of the fitted curves is $1$ for CWB and $0.999$ for CWB with binning. 

\subsubsection{ACWB}

\begin{figure}[ht]
    \centering
    \includegraphics[width=0.5\textwidth, trim=0cm 0.4cm 0cm 0.5cm]{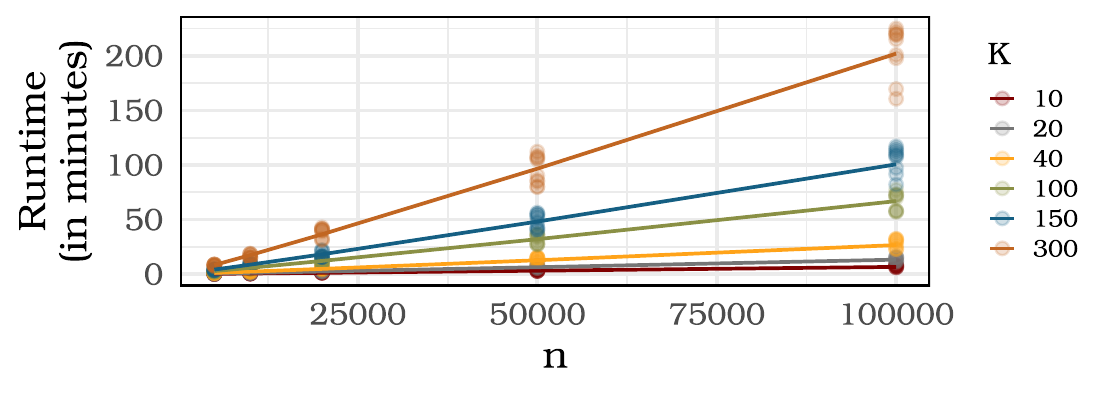}
    \caption{Fitted curves based on computational complexity with fitting time of ACWB used as proxy and response variable. The number of rows $n$ and base learners $K$ are used as input.}
    \label{fig:h4-comp-compl}
\end{figure}

Following our derivations in Section~3.2.2, we expect the runtime of ACWB to scale linear with the number of observations $n$ as well as the number of base learners $K$. Therefore, we model the time as proxy for the computational complexity. The fit confirms our complexity estimates with $\mathcal{O}(K^{0.861}(\bd^{2.165}n^{0.835} + \bd^{2.742}) + 2M^{0.769}K^{1.002}(\bd^{1.904} + d^{1.134}n^{1.087}))$. Figure~\ref{fig:h4-comp-compl} shows the fitted curves. Additionally, a scale parameter $v = 2.273\cdot 10^7$ was estimated to account for different scales of seconds and number of operations. The R-square of the fitted curves is $0.9779$. As shown, ACWB scales efficiently with $\mathcal{O}(n)$ and $\mathcal{O}(K)$. Considering the complexity of HCWB, it is sufficient to know $\mathcal{O}(\text{HCWB}) = a\mathcal{O}(\text{CWB}) + b\mathcal{O}(\text{ACWB}) = \mathcal{O}(\text{ACWB})$. It is worth noting that the empirical and theoretical claims do not show how fast the algorithms are in practice due to the use of early stopping procedures as well as the simplification of just considering numerical features with $d_k = d$ and using a fixed number of iterations. 

\subsection{Hyperband Schedule}\label{app:hypberband-schedule}

Table~\ref{tab:hyperband-schedule} contains the schedule used for all algorithms to tune the HPs.

\begin{table}[ht]
    \centering\footnotesize

\begin{tabular}{r|r|r|r}
\hline\hline
\textbf{Bracket} & \textbf{Stage} & \textbf{Budget} & \textbf{\#HP configurations}\\
\hline\hline
7 & 0 & 39 & 128\\
\hline
7 & 1 & 78 & 64\\
\hline
7 & 2 & 156 & 32\\
\hline
7 & 3 & 312 & 16\\
\hline
7 & 4 & 625 & 8\\
\hline
7 & 5 & 1250 & 4\\
\hline
7 & 6 & 2500 & 2\\
\hline
7 & 7 & 5000 & 1\\
\hline\hline
6 & 0 & 78 & 74\\
\hline
6 & 1 & 156 & 37\\
\hline
6 & 2 & 312 & 18\\
\hline
6 & 3 & 625 & 9\\
\hline
6 & 4 & 1250 & 4\\
\hline
6 & 5 & 2500 & 2\\
\hline
6 & 6 & 5000 & 1\\
\hline\hline
5 & 0 & 156 & 43\\
\hline
5 & 1 & 312 & 21\\
\hline
5 & 2 & 625 & 10\\
\hline
5 & 3 & 1250 & 5\\
\hline
5 & 4 & 2500 & 2\\
\hline
5 & 5 & 5000 & 1\\
\hline\hline
4 & 0 & 312 & 26\\
\hline
4 & 1 & 625 & 13\\
\hline
4 & 2 & 1250 & 6\\
\hline
4 & 3 & 2500 & 3\\
\hline
4 & 4 & 5000 & 1\\
\hline\hline
3 & 0 & 625 & 16\\
\hline
3 & 1 & 1250 & 8\\
\hline
3 & 2 & 2500 & 4\\
\hline
3 & 3 & 5000 & 2\\
\hline\hline
2 & 0 & 1250 & 11\\
\hline
2 & 1 & 2500 & 5\\
\hline
2 & 2 & 5000 & 2\\
\hline\hline
1 & 0 & 2500 & 8\\
\hline
1 & 1 & 5000 & 4\\
\hline\hline
0 & 0 & 5000 & 8\\
\hline\hline
\end{tabular}

    \caption{The schedule used by Hyperband for HP optimization. The total number of tried HP configuration (314) is given by accumulating the number of HP configurations in each bracket at stage 0.}
    \label{tab:hyperband-schedule}
\end{table}

\subsection{EQ1: Figure of comparing \texttt{compboost} with \texttt{mboost}}

\begin{figure}[ht]
    \centering
    \includegraphics[width=\textwidth]{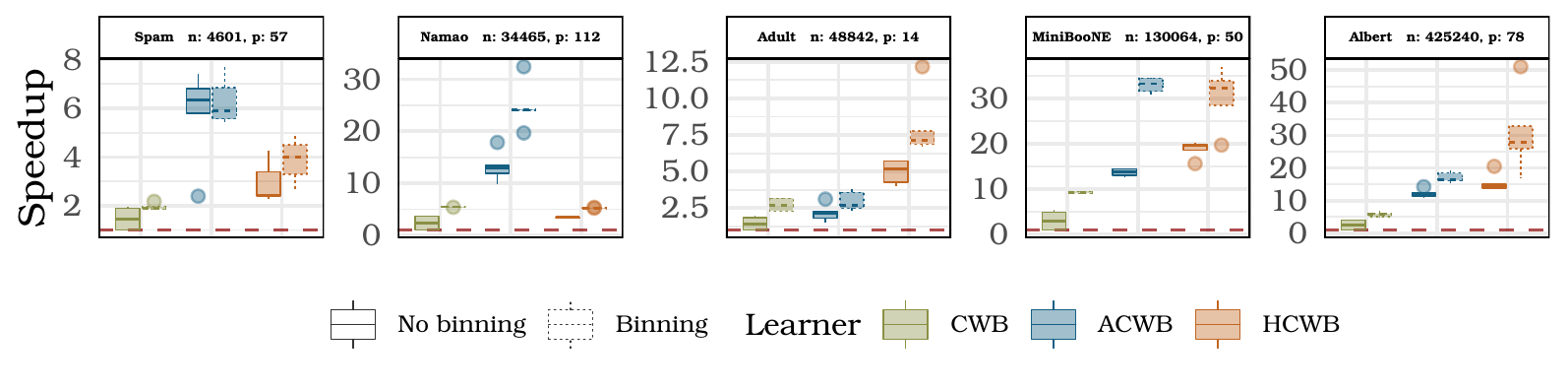}
    \caption{Speedup of our CWB variants compared to the state-of-the-art implementation \texttt{mboost}. Values on the y-axis visualize the speedup calculated as the runtime of \texttt{mboost} divided by the runtime of our implementations.}
    \label{fig:eq0-mboost}
\end{figure}

Figure~\ref{fig:eq0-mboost} visualizes the speedup of CWB/ACWB/HCWB implemented in \texttt{compboost} when compared with the CWB implementation of \texttt{mboost}. It should be noted that there is no data available for the Christine data set since it was not possible to train CWB using \texttt{mboost} for this data set.

\subsection{EQ2: Relative Speedup and AUC Improvement}

Table~\ref{tab:eq1-improvements} shows the speedup of our CWB variants when compared with vanilla CWB. 

\begin{table}[ht]
    \centering\scriptsize

\begin{tabular}{|c|l|r|r|r|r|r|r|r|r|r|r|r|r|}
\hline
\textbf{Learner /} 
  & \multicolumn{2}{c|}{\textbf{Spam}} 
  & \multicolumn{2}{c|}{\textbf{Christine}} 
  & \multicolumn{2}{c|}{\textbf{Namao}} 
  & \multicolumn{2}{c|}{\textbf{Adult}} 
  & \multicolumn{2}{c|}{\textbf{MiniBooNE}} 
  & \multicolumn{2}{c|}{\textbf{Albert}} \\ 
\textbf{Binning} 
  & $\Delta$t & $d$AUC\hspace{-0.1cm} 
  & $\Delta$t & $d$AUC\hspace{-0.1cm} 
  & $\Delta$t & $d$AUC\hspace{-0.1cm} 
  & $\Delta$t & $d$AUC\hspace{-0.1cm} 
  & $\Delta$t & $d$AUC\hspace{-0.1cm} 
  & $\Delta$t & $d$AUC\hspace{-0.1cm} \\ 
\hline\hline
CWB (b)   & 1.014 &  0.000 &  2.003 & -0.002 & 1.482 &  0.000 & 1.363 & 0.000 & 1.866 & 0.000 & 1.375 & 0.000\\
\hline\hline
ACWB (nb) & 3.147 & -0.013 &  9.072 &  0.003 & 3.505 & -0.016 & 1.173 & 0.001 & 2.893 & 0.003 & 2.963 & 0.002\\
\hline
ACWB (b)  & 3.277 & -0.012 & 15.476 & -0.001 & 6.554 & -0.012 & 1.500 & 0.001 & 6.435 & 0.003 & 3.950 & 0.001\\
\hline\hline
HCWB (nb) & 1.451 & -0.005 &  2.008 & -0.009 & 0.940 & -0.003 & 2.582 & 0.001 & 3.936 & 0.002 & 3.388 & 0.001\\
\hline
HCWB (b)  & 1.989 & -0.006 &  1.502 &  0.001 & 1.422 & -0.004 & 3.856 & 0.001 & 6.627 & 0.002 & 7.547 & 0.000\\
\hline
\end{tabular}

    \caption{Relative speedup ($\Delta$t) and AUC improvement ($d$AUC) of the 5 CWB variants compared to CWB (nb). The AUC improvement is calculated as $\text{AUC}_{\text{learner}} - \text{AUC}_{\text{CWB (nb)}}$.}
    \label{tab:eq1-improvements}
\end{table}

\subsection{EQ2: Summary of GLM and ANOVA on the AUC}

Conducting the GLM with the AUC as response contains the task, learner (CWB variant), and a binary variable if binning was applied. Additionally, interactions between binning and the learner are included. The model formula is given by
$$
auc \sim task + binning + optimizer + binning * learner.
$$
Table~\ref{tab:gam-res} and \ref{tab:aov-res} contains the \texttt{R} output of the beta regression and the corresponding ANOVA.

\begin{table}[ht]
    \centering\footnotesize

\begin{tabular}{l|r|r|r|r}
\hline
  & Estimate & Std. Error & z value & Pr(>|z|)\\
\hline
(Intercept) & 3.5672 & 0.0310 & 115.1075 & 0.0000\\
\hline
task168908 & -2.1910 & 0.0297 & -73.7020 & 0.0000\\
\hline
task7592 & -1.1963 & 0.0319 & -37.5123 & 0.0000\\
\hline
task168335 & 0.0456 & 0.0393 & 1.1593 & 0.2463\\
\hline
task189866 & -2.5422 & 0.0294 & -86.5780 & 0.0000\\
\hline
task9977 & 0.3423 & 0.0425 & 8.0538 & 0.0000\\
\hline
binningBinning & -0.0011 & 0.0220 & -0.0510 & 0.9594\\
\hline
learnerACWB & -0.0180 & 0.0220 & -0.8200 & 0.4122\\
\hline
learnerhCWB & -0.0087 & 0.0220 & -0.3952 & 0.6927\\
\hline
binningBinning:learnerACWB & -0.0035 & 0.0311 & -0.1115 & 0.9112\\
\hline
binningBinning:learnerhCWB & 0.0025 & 0.0311 & 0.0796 & 0.9366\\
\hline
\end{tabular}

    \caption{Summary results of the GAM.}
    \label{tab:gam-res}
\end{table}

\begin{table}[ht]
    \centering\footnotesize

\begin{tabular}{l|r|r|r}
\hline
  & df & Chi.sq & p-value\\
\hline
task & 5 & 20785.3355 & 0.0000\\
\hline
binning & 1 & 0.0026 & 0.9594\\
\hline
learner & 2 & 0.6727 & 0.7144\\
\hline
binning:learner & 2 & 0.0370 & 0.9817\\
\hline
\end{tabular}

    \caption{Results of the ANOVA applied on the GLM.}
    \label{tab:aov-res}
\end{table}

\subsection{EQ3: Table of the Renchmark Result}

\begin{table}[ht]
    \centering\tiny
    
\begin{tabular}{|ll|c|c|c|c|c|c|} 
\hline 
& \multirow{2}{*}{\textbf{Algorithm}} & \multicolumn{6}{c|}{\textbf{Data set}} \\ 
 & & \textbf{Spam} & \textbf{Christine} & \textbf{Namao} & \textbf{Adult} & \textbf{MiniBooNE} & \textbf{Albert}
\\ 
\hline\hline 
\multirow{4}{*}{AUC} & ACWB (b) & $0.969 \pm 0.009$ & $0.794 \pm 0.019$ & $0.98 \pm 0.001$ & $0.915 \pm 0.002$ & $0.974 \pm 0.001$ & $0.735 \pm 0.001$\\ 
\cline{2-8}
 & hCWB (b) & $0.969 \pm 0.009$ & $0.797 \pm 0.016$ & $0.978 \pm 0.001$ & $0.916 \pm 0.002$ & $0.975 \pm 0.001$ & $0.735 \pm 0.001$\\ 
\cline{2-8}
 & EBM & $0.986 \pm 0.005$ & $0.814 \pm 0.012$ & $0.994 \pm 0.000$ & $0.928 \pm 0.001$ & $0.978 \pm 0.000$ & $0.747 \pm 0.000$\\ 
\cline{2-8}
 & XGBoost & $0.99 \pm 0.002$ & $0.823 \pm 0.016$ & $0.996 \pm 0.000$ & $0.929 \pm 0.001$ & $0.987 \pm 0.000$ & $0.756 \pm 0.001$\\ 
\hline\hline 
\multirow{4,}{*}{\makecell{Runtime \\ (in hours)}} & ACWB (b) & $2.823 \pm 0.008$ & $17.741 \pm 0.274$ & $3.884 \pm 0.009$ & $1.665 \pm 0.022$ & $7.419 \pm 0.027$ & $24.74 \pm 0.193$\\ 
\cline{2-8}
 & hCWB (b) & $2.832 \pm 0.009$ & $27.361 \pm 0.266$ & $3.978 \pm 0.026$ & $1.511 \pm 0.021$ & $7.476 \pm 0.047$ & $26.249 \pm 0.228$\\ 
\cline{2-8}
 & EBM & $3.711 \pm 0.103$ & $164.228 \pm 0.441$ & $15.668 \pm 0.000$ & $8.172 \pm 0.426$ & $28.183 \pm 0.751$ & $148.927 \pm 0.000$\\ 
\cline{2-8}
 & XGBoost & $0.911 \pm 0.010$ & $6.053 \pm 0.582$ & $3.829 \pm 0.117$ & $3.975 \pm 0.224$ & $15.401 \pm 0.3$ & $96.727 \pm 5.029$\\ 
\hline 
\end{tabular}

    \caption{\footnotesize Average and standard deviation of AUC values as well as runtimes (in hours) for the 5 outer folds of our experiments.}
    \label{tab:perf-values}
\end{table}

\end{appendix}
\newpage
\bibliographystyle{chicago}
\setlength{\bibsep}{0pt plus 0.3ex}
\begin{footnotesize}
\bibliography{references}
\end{footnotesize}